\RequirePackage{amsmath}
\documentclass[12pt]{iopart}
 
\usepackage{amsmath,amssymb}
\usepackage{amsthm}        
\usepackage{graphicx}

\theoremstyle{definition}
\theoremstyle{definition} 
\theoremstyle{definition}\newtheorem{proposition}{Proposition} 
\theoremstyle{definition} 

\begin{document}

\title{Characterisation of circular light rays in a plasma}
\author{Volker Perlick}
\address{
Faculty 1, University of Bremen \\ 28359 Bremen, Germany \\
Email: perlick@uni-bremen.de}



\begin{abstract}
It is the purpose of this paper to give a characterisation
of circular light rays in a plasma on an axially symmetric 
and stationary spacetime. We restrict to the case of an
unmagnetised, pressure-free electron-ion plasma and 
we assume that the plasma shares the symmetry of 
the spacetime. As a main tool we use two potentials,
one for prograde and one for retrograde light rays, whose
critical points are exactly the circular light rays in the plasma.
In the case that the plasma density vanishes, the
corresponding equipotential surfaces reduce to the 
relativistic Von Zeipel cylinders which have been discussed
in many papers since the 1970s. In a plasma, the
gradients of the potentials give the centrifugal and the
Coriolis forces experienced by a light ray, where the
plasma has an influence only on the centrifugal force.
The introduction of these potentials allows us to 
generalise topological methods that have been 
successfully used for proving the existence or
non-existence of circular vacuum light rays to the 
plasma case. The general results are illustrated with
examples on Minkowski, Schwarzschild, Kerr and NUT
spacetimes.
\end{abstract}

\section{Introduction}

If gravitating objects are sufficiently compact, the light-bending
effect of their gravitational fields may be so strong that some 
light rays are circular. Examples for such objects, which are sometimes
called \emph{ultracompact}, are black holes and wormholes and
there are also speculations that ultracompact stars might exist.
The existence of circular light rays has important consequences
for the light-bending effects in general. In particular, circular 
light rays that are unstable against perturbations in the radial 
direction are associated with light rays that asymptotically spiral 
towards them, giving rise to infinite bending angles and to
the formation of a \emph{shadow}. For reviews on the latter
we refer to Cunha and Herdeiro \cite{CunhaHerdeiro2018}
and to Perlick and Tsupko \cite{PerlickTsupko2022}.

When discussing the light bending of gravitating objects, it is 
justified in most cases to treat the light rays as lightlike 
geodesics of the spacetime metric, i.e., to disregard any 
influence of a medium. For some applications, however, 
this influence might be non-negligible. This is true, in
particular, in the radio regime if the medium is a 
sufficiently dense plasma. E.g., for a light ray that crosses 
the ionosphere of our Earth or the Solar corona, it is known 
that there is a measurable influence of the medium on the 
travel time and on the spatial path of a light ray unless the 
frequency is much bigger than 10 MHz or 100 MHz, respectively.
It is not unplausible to assume that near a black hole the 
plasma density might be even higher, i.e., to assume that 
near a black hole light rays of even higher frequencies could 
be influenced by a plasma in a non-negligible way. This gives 
a good motivation for investigating the influence of a 
plasma on the lensing features of an ultracompact object.

Gravitational light deflection on a general-relativistic
spacetime in the presence of a plasma has been investigated
by many authors. The first papers on this subject date back
to the 1960s when the Shapiro time delay of light rays 
(in the radio regime) that pass through the Solar corona
was at the centre of interest. Muhleman et al. 
\cite{MuhlemanJohnston1966,MuhlemanEkersFomalont1970}
calculated, on the basis of the linearised Schwarzschild metric, 
the time delay and the bending angle of such light rays.
Light bending in a plasma, without the weak-field approximation,
on the Schwarzschild spacetime and in the equatorial plane of 
a Kerr metric, was first calculated by Perlick \cite{Perlick2000b}. 
This was followed by many other papers on light bending by a 
gravitational field in the presence of a plasma. Here, in 
particular the work by Bisnovatyi-Kogan and Tsupko 
\cite{BisnovatyiTsupko2009,BisnovatyiTsupko2010}
has to be mentioned, which considerably pushed the subject, 
and two papers by Perlick and Tsupko 
\cite{PerlickTsupko2017,PerlickTsupko2024} where 
the influence of a plasma on lensing by a Kerr black hole  
was discussed in detail.  In all these papers, the simplest model 
of a plasma was used, namely a non-magnetised pressure-less
electron-ion plasma. 

It is the purpose of the present paper to present a mathematical
formalism that allows to  investigate the existence of circular 
light rays in the presence of a plasma. As outlined in the 
preceding paragraphs, this is motivated by the two facts that
the existence of circular light rays is of crucial relevance for 
the light bending and that the influence of a plasma might
be non-negligible in some cases of interest. To that end, we
restrict to axially symmetric and stationary spacetimes and we
assume that the plasma shares these symmetries. Moreover, 
as in the works quoted above, we restrict to the simple case
of a non-manetised pressure-less electron-ion plasma. As a 
main tool, we define two potentials, one for prograde and one
for retrograde light rays, in such a way that the critical points
of these potentials give the locations of circular light rays. If
there is no plasma, the corresponding equipotential surfaces
reduce to the \emph{relativistic Von Zeipel cylinders} which
were introcuded by Marek Abramowicz, originally for axially 
symmetric and static spacetimes,  already in the 1970s and
have been discussed in numerous papers, see in particular 
Abramowicz \cite{Abramowicz1971,Abramowicz1974,Abramowicz1990}.
These equipotential surfaces have many interesting properties.
What is most important for our purpose is the fact that they
give us the circular light rays which are located exactly where
the potentials have critical points. The generalised potentials that
will be introduced in this paper have the same crucial property 
in the plasma case. 

With the help of these potentials we can visualise
the behaviour of light rays in an instructive way: If a light ray 
is sent tangent to a circle, the gradient of the corresponding
potential points in the direction in which this light ray is
deflected from this circle. The corresponding ``force'' acting
on the light ray can be interpreted as the sum of a 
``centrifugal force'' and a ``Coriolis force''. Interestingly, 
we will see that only the centrifugal force is influenced by 
the plasma. However, as already mentioned it is most 
important that  the location of circular light rays is given 
by the critical points of the corresponding potential. In 
addition to just visualising
the light-deflection properties in a suggestive way, this 
mathematical construction allows us to generalise theorems 
on the existence of circular light rays that have been 
established with topological methods, using the Brouwer 
deegree of appropriate maps, by Cunha et al. 
\cite{CunhaBertiHerdeiro2017,CunhaHerdeiro2020,CunhaHerdeiroNovo2024},
to the plasma case.

It should be emphasised that throughout the paper Einstein's field
equation is not used, so the general results apply to the case that 
the plasma is self-gravitating and equally well to the case that the 
gravitational field of the plasma is neglected. The latter case is,
probably, more interesting in view of applications to astrophysics,
so in our examples we consider only vacuum solutions to Einstein's 
field equation, namely Minkowski, Schwarzschild, Kerr and NUT 
spacetimes.  

In Section \ref{sec:plasma} we briefly review the Hamiltonian
formalism for light rays in a non-magnetised pressure-less 
electron-ion plasma on an unspecified general-relativistic 
spacetime. In Section \ref{sec:axistat} we specialise to the
axially symmetric and stationary case. In particular, we 
characterise, in Proposition \ref{prop:circular}, light rays 
that are tangent to a circle at one point. This proposition
is crucial for the rest of the paper. On the basis of this 
proposition, we introduce in Section \ref{sec:defR} two 
potentials $\mathcal{R}{}_{\pm}$. 
Proposition \ref{prop:circular} makes sure that these
potentials determine the centrifugal and the Coriolis force
acting on light rays in the plasma and that, in particular,
the critical points determine the locations of circular
light rays. In Section \ref{sec:prop} we discuss the 
properties of these two potentials, in particular their
limiting behaviour if a horizon, a regular axis or an asymptotic
flat end is approached. In Section \ref{sec:theorems}
we point out how the preceding results allow to 
generalise the above-quoted work of Cunha et al. 
to the plasma case. Finally, in Sections \ref{sec:reg}
and \ref{sec:blackh} we illustrate the general results 
with several examples.      
      
Throughout, we use Einstein's summation convention for greek 
indices $\mu, \nu , \dots$ that take values $0,1,2,3$. When 
specialising to axially symmetric and static spacetimes, we will 
occasionally use Einstein's summation convention for upper case
latin indices $A, B, \dots$ that take values $0,1$ and for lower
case latin indices $i,j, \dots$ that take values $2,3$. We use
units making the vacuum speed of light $c$ qual to 1 and 
we assume that the spacetime has signature $(-+++)$.

\section{Light rays in a plasma}\label{sec:plasma}

Light rays in a non-magnetised and pressure-less electron-ion 
plasma on a general-relativistic spacetime with metric 
$g_{\mu \nu}$ are the solutions of Hamilton's equations
\begin{equation}
\dot{x}{}^{\mu} = \dfrac{\partial\mathcal{H}}{\partial p_{\mu}} \, ,
\, \quad
\dot{p}{}^{\mu} = -\dfrac{\partial \mathcal{H}}{\partial x_{\mu}} \, ,
\, \quad
\mathcal{H} = 0
\label{eq:Hamgen}
\end{equation}
with the Hamiltonian
\begin{equation}
\mathcal{H} (x,p) = \dfrac{1}{2} \Big(
g^{\rho \sigma} (x) p_{\rho} p_{\sigma} + \omega _p (x) ^2
\Big)     \, .
\label{eq:Ham}
\end{equation}
Here $\omega _p$ is a non-negative function of the spacetime 
coordinates with the dimension of a frequency. It is known as the 
\emph{plasma frequency}. Its square $\omega _p ^2$ equals, 
up to a constant factor that depends on the system 
of units used, the number density of electrons. We will
refer to $\omega _p ^2$ as to the \emph{plasma density} 
in the following.
 
The overdot denotes the derivative with respect to a curve parameter $s$.
This parameter has no particular physical meaning. It is \emph{not}
an affine parameter, i.e., the equations are \emph{not} invariant
under affine changes of this parameter, unless $\omega _p = 0$. 

For a derivation from Maxwell's equations of this Hamiltonian 
formalism for light rays in a plasma on a curved spacetime we 
refer to Breuer and Ehlers \cite{BreuerEhlers1980} who even 
treated the case of a magnetised plasma. For a non-magnetised 
plasma a derivation can also be found in a book by 
Perlick \cite{Perlick2000b}. A non-magnetised pressure-less plasma 
belongs to the class of isotropic dispersive media which have 
been treated first in a book by Synge \cite{Synge1960}. 

Written out explicitly, the equations (\ref{eq:Hamgen}) with
the Hamiltonian (\ref{eq:Ham}) read
\begin{equation}
\dot{x}^{\mu} = g^{\mu \sigma} p_{\sigma} \,,
\label{eq:Hamx}
\end{equation}
\begin{equation}
\dot{p}_{\mu} = - \dfrac{1}{2} \Bigg(
\dfrac{\partial g^{\rho \sigma}}{\partial  x^{\mu}}
p_{\rho} p_{\sigma} + \dfrac{\partial \omega _p ^2}{\partial x^{\mu}}
\Bigg)\, ,
\label{eq:Hamp}
\end{equation}
\begin{equation}
\mathcal{H} = \dfrac{1}{2} \Big(
g^{\rho \sigma} p_{\rho} p_{\sigma} + \omega _p ^2 \Big)
= 0 \, .
\label{eq:disp}
\end{equation}
Eq. (\ref{eq:Hamx}) is equivalent to 
\begin{equation}
p_{\nu} = g_{\nu \rho } \dot{x}^{\mu}  \, ,
\label{eq:Ham1inv}
\end{equation}
so (\ref{eq:disp}) can be rewritten as
\begin{equation}
\mathcal{H} =
\dfrac{1}{2} \Big( p_{\rho} \dot{x}{}^{\rho} 
+ \omega _p ^2 \Big)
=0   
\label{eq:Ham2}
\end{equation}
or
\begin{equation}
\mathcal{H} =
\dfrac{1}{2} \Big( g_{\rho \sigma} \dot{x}{}^{\rho} \dot{x}{}^{\sigma} 
+ \omega _p ^2 \Big) 
=0 \, .    
\label{eq:Ham3}
\end{equation}
From (\ref{eq:Ham3}) we read that light rays in a plasma are 
\emph{timelike} curves. More precisely, they are \emph{timelike 
geodesics} of the conformally rescaled metric $\omega _p ^2 \, 
g_{\mu \nu}$, as can be easily verified. The parameter $s$ is
proper time with respect to this (unphysical) metric.  

If we choose an observer field, i.e., a timelike vector field $U^{\mu}$
with $g_{\mu \nu} U^{\mu} U^{\nu} = - 1$, we can assign a 
frequency 
\begin{equation}
\omega (s) = -p_{\mu} (s) \, U^{\mu} \big ( x (s) \big)
\label{eq:omega}
\end{equation}
to every light ray. The minus sign makes sure that $\omega$ is positive 
for a light ray that is future-oriented with respect to $U^{\mu}$.
In vacuum, the frequency of a light ray  has no influence on its path.
In a plasma, however, the dispersion relation (\ref{eq:disp}) is 
inhomogeneous (i.e., not invariant under multiplication of $p_{\mu}$
with a non-zero factor), which implies that light rays that start in the same 
spatial direction with different frequencies have different trajectories.
This inhomogeneity is the defining property of a \emph{dispersive
medium}, see e.g. Perlick \cite{Perlick2000b} for a detailed discussion.

\section{The axially symmetric and stationary case}
\label{sec:axistat}

We now specify to the case that the metric is axially symmetric and 
stationary and that the plasma density shares this symmetry. 
This means that the metric can be written in coordinates 
$(x^0 = t , x^1 = \varphi , x^2 , x^3)$ such that 
\begin{equation}
\dfrac{\partial g_{\mu \nu}}{\partial x^A} = 0 \, , \quad
\dfrac{\partial \omega _p ^2}{\partial x^A} = 0 \, .
\end{equation}
Here and in the following we use the summation convention 
for upper-case latin indices $A, B, \dots$$\, =0,1$ and for 
lower-case latin indices $i,j, \dots$$\, = 2,3$. The $x^i$ could be, e.g., 
spherical polar coordinates $(r, \vartheta)$ or cylindrical polar 
coordinates $(\rho , z)$. As before, greek indices take values
$0,1,2,3$. We assume that the metric coefficients $g_{AB}$
and the plasma density $\omega _p ^2$ are smooth ($C^2$ 
would do) on the domain under consideration. 

We will assume in the following that the coordinate function
$\varphi$ is $2 \pi$-periodic,
\begin{equation}
(t , \varphi , x^2, x^3) \equiv (t, \varphi + 2 \pi , x^2 , x^3 ) \, ,
\label{eq:period}
\end{equation}
and that the two-dimensional surfaces $x^i = \mathrm{constant}$ 
are timelike, i.e.
\begin{equation}
g_{ \varphi \varphi} g_{tt} - g_{t \varphi}^2 < 0
\, .
\label{eq:det}
\end{equation}
The first condition means that $\varphi$ can be interpreted as
an angular coordinate, the second condition means that there
is a linear combination of the Killing vector fields $\partial _t$
and $\partial _{\varphi}$ that is timelike. Both assumptions 
are necessary to justify calling the spacetime ``axially symmetric 
and stationary''. Note that in a black-hole spacetime the condition
(\ref{eq:det}) is satisfied in the domain of outer cmmunication,
i.e., outside the horizon, while the left-hand side of (\ref{eq:det})
is equal to zero on the horizon and positive inside.

As the spacetime metric $( g _{\mu \nu}) $ has signature $(-+++)$,
(\ref{eq:det}) implies that the $(2 \times 2)$-matrix $(g^{ij})$ is 
positive definite. This will be relevant for the following considerations.

Note that we do \emph{not} require the mixed metric components 
$g_{iA}$ to be zero. If the metric is invariant under a transformation
$(t, \varphi ) \mapsto (- t , - \varphi)$, the $g_{iA}$ must be zero.
It is indeed true that such a choice of coordinates is possible for most 
axially symmetric and stationary spacetimes with relevance to physics. 
However,  even in these cases it is nonetheless meaningful to allow for 
another choice of coordinates where $g_{i A} \neq 0$. An example 
is the Schwarzschild spacetime in Eddington-Finkelstein or 
Painlev{\'e}-Gullstrand coordinates.

Also note that we do \emph{not} require the spacetime to admit
a horizontal plane, i.e., we do \emph{not} require that it is possible 
to choose cylindrical polar coordinates $(\rho, z)$ such that the 
metric coefficients are invariant under a transformation $z \mapsto -z$.

Moreover, we will allow the metric functions $g_{tt}$ and 
$g_{\varphi \varphi}$ to have either sign, i.e., we will allow for the
presence of an ergoregion, where $g_{tt} > 0$, and of a
causality-violating region, where $g_{\varphi \varphi} < 0$.
An ergoregion occurs, e.g., in the domain of outer communication
of a Kerr black hole while a causality-violating region occurs, e.g., 
in the domain of outer communication of a NUT black hole.

The following proposition, which is at the basis of all that follows, 
characterises light rays that are tangent, at one point, to a surface
${x}{}^i = \mathrm{constant}$. Obviously, this applies in
particular to circular light rays which, by definition, are light rays 
with $\dot{\varphi} \neq 0$ that are completely contained in a 
surface ${x}{}^i = \mathrm{constant}$. 
  
\begin{proposition}
Let $x^{\mu}(s)$ be a solution to Hamilton's equations in an
axially symmetric and stationary spacetime. Then, at points 
where $\dot{x}{}^i = 0$, the following equations are true.
\begin{equation}
p_{\varphi}
= 
\dfrac{g_{t \varphi}}{g_{tt}} \, p_t \mp
\dfrac{1}{g_{tt}} \, \sqrt{g_{t \varphi} ^2 -g_{tt} g_{\varphi \varphi}}
\,
\sqrt{p_t^2+ \omega _p^2 g_{tt}}
\label{eq:pphiR} 
\end{equation}

\begin{equation}
\dot{\varphi}
= 
\pm \,  
\dfrac{
\sqrt{p_t ^2+ \omega _p^2 g_{tt}}
}{
\sqrt{g_{t \varphi} ^2 -g_{tt} g_{\varphi \varphi}}
}
\,
\label{eq:dotphi} 
\end{equation}

\begin{equation}
\dot{t}
= 
\dfrac{p_t}{g_{tt}} 
\mp
\dfrac{g_{t \varphi}}{g_{tt}} \,
\dfrac{
\sqrt{p_t^2+ \omega _p^2 g_{tt}}
}{
\sqrt{g_{t \varphi} ^2 -g_{tt} g_{\varphi \varphi}}
}
\, 
\label{eq:dott} 
\end{equation}

\vspace{0.3cm}

\begin{eqnarray}
\, \hspace{-1cm}
\ddot{x}{}^{\mu}
= 
\, - \, 
\dfrac{ 
\sqrt{p_t^2+ \omega _p^2 g_{tt}}
}{
\sqrt{
g_{t \varphi} ^2  - g_{tt} g_{\varphi \varphi}
}
}
\, 
g^{\mu i}
\,
\dfrac{\partial}{\partial x^i}
\left(
\mp
\dfrac{g_{t \varphi}}{g_{tt}} \, p_t 
+
\dfrac{1}{g_{tt}}
\sqrt{ g_{t \varphi} ^2 - g_{tt} g_{\varphi \varphi}}
\,
\sqrt{p_t^2+ \omega _p^2 g_{tt}}
\; 
\right)
\label{eq:ddx} 
\end{eqnarray}
The upper sign is for prograde light rays ($\dot{\varphi} > 0$) 
and the lower sign for retrograde ones ($\dot{\varphi} > 0$). 
\label{prop:circular}
\end{proposition}

\hspace{0.2cm}

The proof of this proposition is given in the appendix. It is a 
straight-forward generalisation from the case $\omega _p =0$. 
For the latter, cf. e.g. Cunha et al. \cite{CunhaBertiHerdeiro2017}.

Proposition \ref{prop:circular} gives us all relevant information
where in phase space light rays with $\dot{x}{}^i = 0$ can 
exist. Firstly, we see that (\ref{eq:det})
must be satisfied. We have already mentioned that this 
condition assures that the two-space $x^i = \mathrm{constant}$
contains a timelike vector. In a plasma, where light rays are 
timelike, this condition is obviously necessary for the existence
of a light ray with $\dot{x}{}^i =0$, and this is confirmed by
the occurrence of the $\sqrt{g_{t \varphi}^2 - 
g_{tt} g_{\varphi \varphi}}$ terms in (\ref{eq:pphiR}), (\ref{eq:dotphi}),
(\ref{eq:dott}) and (\ref{eq:ddx}). One might argue that in the vacuum 
case, where the light rays are lightlike, we should also allow the 
limiting case that the surface $x^i = \mathrm{constant}$ is
lightlike, i.e., $g_{t \varphi}^2- g_{tt} g_{\varphi \varphi}
= 0$.  It is indeed true that vacuum light rays can be contained
in such a surface, but this would be the generators of a horizon and
we do not want them to be considered as circular light rays. So
we require, also in vacuum, the strict inequality (\ref{eq:det}).

Secondly, we read from Proposition \ref{prop:circular} that we should
restrict to the region in phase space where 
\begin{equation}
p_t^2+ \omega _p ^2 g_{tt} > 0
\, .
\label{eq:bound0}
\end{equation}
Again, one might argue about the limiting case that the left-hand
side of (\ref{eq:bound0}) is equal to zero, but then (\ref{eq:dotphi})
can hold only with $\dot{\varphi} =0$ and we do not consider 
such light rays as circular. To be sure, it is obvious and well known 
that light rays with this property do exist in a plasma: In a region
with $\omega _p \neq 0$ and $g_{tt} < 0$ there is always a 
value of $|p_t|$ such that the left-hand side of (\ref{eq:bound0}) 
is zero. Light rays with this frequency are tangent to a $t$-line, i.e.,
they ``stand still''. This happens, e.g., at the boundary of the 
Earth's ionosphere for $|p_t| \approx$ 10 MHz. A light ray with 
a lower frequency is reflected from the boundary of the 
ionosphere. As we do not consider a light ray as ``circular'' 
if it stands still, we require that (\ref{eq:bound0})  holds 
with the strict ineqality sign. 

Thirdly, the occurrence of factors of $g_{tt}$ in the denominators
of (\ref{eq:pphiR}), (\ref{eq:dott}) and (\ref{eq:ddx}) requires 
to investigate what happens if $g_{tt} = 0$. If we multiply 
in (\ref{eq:pphiR}) and (\ref{eq:dott}) both sides with $g_{tt}$ 
and use the fact that $p_{\varphi}$ and $\dot{t}$ must be finite, 
we see that in the limit $g_{tt} \rightarrow 0$ the condition 
$g_{t \varphi} p_t = 
\pm | g_{t \varphi} p_t |$ must hold. As $g_{tt} =0$ requires,
by (\ref{eq:det}) and (\ref{eq:bound0}), that $g_{t \varphi} \neq 0$ 
and $p_t \neq 0$, this condition holds for only one of the two signs,
i.e., at points where $g_{tt} = 0$ only a prograde or a retrograde
circular light ray can exist, but not both. This gives us the following
third restriction on the phase space:
\begin{equation}
g_{tt}^2 + p_t^2 
\Big(   \pm g_{t \varphi} + \big| g_{t \varphi} \big| \, \Big) ^2 > 0 \, ,  
\label{eq:pm}
\end{equation}
where the upper sign gives the restriction for prograde and the 
lower sign for retrograde light rays.

\section{Definition of the potentials $R_{\pm}$}
\label{sec:defR}

As a plasma is a dispersive medium, the propagation of light rays
depends on their frequencies. In a stationary spacetime, where $p_t$
is a constant of motion, it is convenient to consider all light rays
with fixed $p_t$. Henceforth, we write
\begin{equation}
p_t = - \omega _0 
\label{eq:omega0}
\end{equation}
and it is our goal to characterise all circular light rays with fixed
$\omega _0$. Note that in an asymptotically flat spacetime, 
where 
\begin{equation}
U^{\mu} = \dfrac{1}{\sqrt{-g_{tt}}} \,  \delta ^{\mu}_t
\longrightarrow
\delta ^{\mu}_t
\label{eq:Umu}
\end{equation}
at infinity, $\omega _0$ is the frequency (\ref{eq:omega})
with respect to this observer field $U ^ {\mu}$ at infinity 
(provided that the light ray under consideration reaches infinity).
We have chosen the sign in (\ref{eq:omega0}) such that in
regions where $\partial _t$ is timelike $\omega _0$ is positive
for light rays that are future-oriented with respect to 
$\partial _t$.

For any fixed $\omega _0 \ge 0$, we now introduce the following 
potentials $\mathcal{R}{}_+$ and $\mathcal{R}{}_-$, which
is motivated by Proposition \ref{prop:circular}.
\begin{equation}
\mathcal{R} _{\pm} = 
\pm \dfrac{g_{t \varphi}}{g_{tt}} \, \omega _0
+
\dfrac{1}{g_{tt}} \,
\sqrt{g_{t \varphi} ^2  - g_{tt} g_{\varphi \varphi}}
\,
\sqrt{\omega _0^2+ \omega _p^2 g_{tt}}
\, .
\label{eq:R}
\end{equation}
Restricting to values $\omega _0 \ge 0$ is justified because the
potentials for $- \omega _0$ are the same as for $\omega _0$, 
just interchanged. As we will outline in Section \ref{subsec:sign} below,
this convention has the consequence that inside an ergoregion the
upper sign in our equations corresponds to a future-oriented parametrisation
of the light rays and the lower sign to a past-oriented one, or vice versa.
Note that in a plasma it is indeed possible that a circular light ray with 
$\omega _0 = 0$ exists: In vacuum, (\ref{eq:bound0}) excludes the 
case $p_t =0$; in a plasma, however, $\omega _p ^2 \, g_{tt}$ is 
positive in an ergoregion, so the case $p_t = 0$ is not forbidden by (\ref{eq:bound0}). Strictly speaking, $\mathcal{R}{}_{\pm}$ should
carry an index $\omega _0$. We do not do this, to ease notation,
so we have to keep in mind that $\mathcal{R}{}_{\pm}$ is to be
considered for a fixed $\omega _0 \ge 0$.

With the help of these potentials $\mathcal{R}{}_{\pm}$, 
(\ref{eq:pphiR}) can be rewritten as
\begin{equation}
p_{\varphi} = \mp \mathcal{R}{}_{\pm}
\end{equation}
and (\ref{eq:ddx}) can be rewritten, for $\mu = j$, as
\begin{equation}
\ddot{x}{}^{j} =\, - \, 
\dfrac{
\sqrt{\omega _0^2 + \omega _p^2 g_{tt}}
}{
\sqrt{
g_{t \varphi}^2 - g_{tt} g_{\varphi \varphi}
}
}
\,
g^{j i } \dfrac{\partial \mathcal{R}_{\pm}}{\partial x^i}
\, .
\label{eq:force}
\end{equation}
As the $2 \times 2$ matrix $(g^{ij})$ is invertible (and 
even positive definite), this equation shows that a light 
ray that is sent tangential to a circle $x^i = 
\mathrm{constant}$ is experiencing a ``force''
that is a negative multiple of the gradient of 
$\mathcal{R}{}_{\pm}$. In particular,
it implies that a circular light ray exists at $x^i
= x^i_0$ if and only if $\mathcal{R}{}_{\pm}$
has a critical point at $x^i = x^i_0$.
Defining the potential such that the direction 
of the force is opposite to the gradient of the potential
is the usual convention in mechanics which implies 
that a stable critical point corresponds to  a local minimum
of the potential.  

Drawing the surfaces $\mathcal{R}_{\pm} = \mathrm{constant}$
illustrates the way in which light rays are deflected and, in 
particular, where circular light rays exist. These equipotential
surfaces generalise the \emph{relativistic Von Zeipel
cylinders} which were introduced by 
Abramowicz \cite{Abramowicz1971,Abramowicz1974} already in
the 1970s  and have been discussed in many papers. To
make the connection, we specify $\mathcal{R}{}_{\pm}$
to the case that $g_{t \varphi} = 0$ and $\omega _p = 0$.
By (\ref{eq:det}), $g_{t \varphi} = 0 $ implies 
$g_{tt} g_{\varphi \varphi} <0$ and, if we exclude 
causality violation, $g_{tt} < 0$, hence 
\begin{equation}
\mathcal{R}_+ = \mathcal{R}_- =
- \sqrt{- \dfrac{g_{\varphi \varphi}}{g_{tt}}} \, \omega _0
\label{eq:Rstatic}
\end{equation}
which is exactly the potential whose equipotential surfaces
have been called the \emph{relativistic Von Zeipel cylinders}
in axisymmetric and static spacetimes, see e.g. 
Abramowicz \cite{Abramowicz1990}.
(For vacuum light rays, these equipotential surfaces are, of
course, independent of the frequency constant $\omega _0$.)  
In flat spacetime, where we can choose cylindrical polar
coordinates such that $g_{tt} = -1$ and $g_{\varphi \varphi} =
\rho ^2$, these equipotential surfaces are indeed straight 
cylinders in three-dimensional space, i.e., they are represented
by vertical lines in the $(x^2,x^3)=(\rho, z )$ half-plane. In 
curved spacetimes, however, they do not always have the 
topology of a cylinder. Therefore, the name should be taken 
with a grain of salt. 

As shown by Abramowicz \cite{Abramowicz1990},
particles in \emph{timelike}
motion along a circle $x^i = \mathrm{constant}$
experience a \emph{centrifugal force} that points in the 
direction of the negative gradient of the potential (\ref{eq:Rstatic}),
with a prefactor that depends on the velocity. The potential 
itself is actually independent of the velocity, so the interpretation
as a potential for the centrifugal force is also valid for particles
that move at the speed of light (although in this case the prefactor
becomes infinite).  In the general case, allowing for a non-zero 
$g_{t \varphi}$ and a non-zero $\omega _p$, the first term on
the right-hand side of (\ref{eq:R}) can be viewed as a potential
for the Coriolis force and the second one as a potential for the
centrifugal force. Note that the plasma only influences the 
centrifugal force. For
a discussion of Von Zeipel cylinders in the case that $\omega _p = 0$
but $g_{t \varphi} \neq 0$ we refer to Jefremov and Perlick 
\cite{JefremovPerlick2016} where
two possible definitions are compared. One of them corresponds
to our potential (\ref{eq:R}) specified to $\omega _p = 0$, the
other one, denoted $\tilde{\mathcal{R}}$ in \cite{JefremovPerlick2016}, 
is of no relevance for the present paper.

\section{Properties of the potentials $\mathcal{R}{}_{\pm}$}  
\label{sec:prop}

\subsection{Domain of definition of the potentials $\mathcal{R}{}_{\pm}$}
\label{subsec:dom}

From the discussion at the end of Section \ref{sec:axistat} we know 
that circular light rays can exist only where (\ref{eq:det}), 
(\ref{eq:bound0}) and (\ref{eq:pm}) hold. Correspondingly, we
consider the potential $\mathcal{R}{}_{\pm}$, for fixed 
$\omega _0 \ge 0$, on the open subset of the 
$(x^2,x^3)$-plane where
\begin{equation}
g_{t \varphi}^2 - g_{tt} g_{\varphi \varphi} > 0
\, ,
\label{eq:det2}
\end{equation}
\begin{equation}
g_{tt}^2 + \omega _0 ^2 \Big( \pm 
g_{t \varphi} + \big|  g_{t \varphi} \big| \,  \Big) ^2 > 0
\, ,
\label{eq:pm2}
\end{equation}
\begin{equation}
\omega _0 ^2 + \omega _p^2 \, g_{t t} > 0
\, .
\label{eq:bound}
\end{equation}
If this set is not connected, we have to consider the potential
$\mathcal{R}{}_{\pm}$ on each connected component
separately. Let us choose one of these connected components, 
denoted $\mathcal{V}{}_{\pm}$. In the following we will discuss
the various possibilities for $\mathcal{V}{}_{\pm}$. We first 
observe that the open set where (\ref{eq:det2}) 
holds need not be connected. We denote the connected component
that contains our chosen $\mathcal{V}{}_{\pm}$ by the letter
$\mathcal{U}$. We call the open subset of $\mathcal{U}$
where (\ref{eq:pm2}) holds $\mathcal{U}{}_{\pm}$.

If $g_{tt} < 0$ on all of $\mathcal{U}$, (\ref{eq:pm2}) gives
no further restriction, hence $\mathcal{U}{}_+ = \mathcal{U}{}_-
=\mathcal{U}$. If $-\omega _p ^2 \, g_{tt}$ is bounded on
$\mathcal{U}$, i.e. if 
\begin{equation}
\mathrm{sup} (-\omega _p^2 \, g_{tt} ) 
< \infty \, ,
\label{eq:omegapbound}
\end{equation}
$\mathcal{R}{}_+$ and $\mathcal{R}{}_-$ are defined on
all of $\mathcal{U}$ if the frequency constant $\omega _0$
has been chosen such that $\omega _0^2 > 
\mathrm{sup} (-\omega _p^2 \, g_{tt} )$, i.e.
$\mathcal{V}{}_+ = \mathcal{V}{}_- = \mathcal{U}$. For
smaller values of $\omega _0$, the sets $\mathcal{V}{}_+$
and $\mathcal{V}{}_-$ are further restricted; on the boundary 
of these restricted domains light rays are tangent to 
a $t$-line, i.e., they ``stand still''.

If $g_{tt} > 0$ on all of $\mathcal{U}$, it is again true 
that (\ref{eq:pm2}) gives no further restriction, hence 
$\mathcal{U}{}_+ = \mathcal{U}{}_-
=\mathcal{U}$, hence $\mathcal{V}{}_+ = \mathcal{V}{}_-
= \mathcal{U}$ for all $\omega _0 > 0$ and in the case 
that $\omega _p$ is bounded away from zero 
on $\mathcal{U}$ even for $\omega _0 = 0$. 

The situation is more complicated if $g_{tt}$ takes positive
and negative values on $\mathcal{U}$. In this case we
decompose $\mathcal{U}$ into the sets
\begin{equation}
\mathcal{U}^{\mathrm{out}} =
\big\{ (x^2, x^3 ) \in \mathcal{U} \, \big| \, 
g_{tt} (x^2 , x^3 ) < 0 \big\} \, , 
\end{equation}
\begin{equation}
\mathcal{U}^{\mathrm{erg}} =
\big\{ (x^2, x^3 ) \in \mathcal{U} \, \big| \, 
g_{tt} (x^2 , x^3 ) > 0 \big\} \, , 
\end{equation}
\begin{equation}
\mathcal{S} =
\big\{ (x^2, x^3 ) \in \mathcal{U} \, \big| \, 
g_{tt} (x^2 , x^3 ) = 0 \big\} \, . 
\end{equation}
$\mathcal{U}^{\mathrm{out}}$ and $\mathcal{U}^{\mathrm{erg}}$
are open while $\mathcal{S}$ is closed in $\mathcal{U}$. If $\mathcal{U}$
is the domain of outer communication of a Kerr black hole, 
$\mathcal{U}^{\mathrm{erg}}$ is the ergoregion, 
$\mathcal{U}^{\mathrm{out}}$
is the region outside the ergoregion, and $\mathcal{S}$ is the 
boundary of the ergoregion in $\mathcal{U}$. 

Because of (\ref{eq:det2}) $g_{t \varphi}$
must be non-zero on $\mathcal{S}$. 
Let us assume that $g_{t \varphi}
< 0$ on $\mathcal{S}$. Then, by (\ref{eq:pm2}), $\mathcal{U}{}_+
= \mathcal{U}$ and $\mathcal{U}_-= \mathcal{U}^{\mathrm{out}}
\cup \mathcal{U}^{\mathrm{erg}}$. If (\ref{eq:omegapbound})
holds on $\mathcal{U}$ the potential $\mathcal{R}{}_+$ is defined
on the domain $\mathcal{V}{}_+= \mathcal{U}$ for all 
$\omega _0 > \mathrm{sup} (-\omega _p^2g_{tt})$. 
For $\mathcal{R}{}_-$ we have to 
consider the domains $\mathcal{U}^{\mathrm{out}}$ and 
$\mathcal{U}^{\mathrm{erg}}$ separately. If 
(\ref{eq:omegapbound}) holds on  $\mathcal{U}^{\mathrm{out}}$,
then $\mathcal{R}{}_-$ is defined on all of $\mathcal{U}^{\mathrm{out}}$
for all $\omega _0 > \mathrm{sup} (-\omega _p^2g_{tt})$,
so for these values of $\omega _0$ 
one possibility for our chosen domain of definition is 
$\mathcal{V}{}_- = \mathcal{U}^{\mathrm{out}}$. On 
$\mathcal{U}^{\mathrm{erg}}$ the potential $\mathcal{R}{}_-$ 
is defined for all $\omega _ 0 > 0$ and in the case that
$\omega _p$ is bounded away from zero on this domain 
even for $\omega _0 = 0$, 
so for these values of $\omega _0$ our chosen domain of 
definition can be $\mathcal{V}{}_- = \mathcal{U}^{\mathrm{erg}}$.
If $\mathcal{S}$ is approached from $\mathcal{U}{}_{\mathrm{out}}$,
the potential $\mathcal{R}{}_-$ goes to $- \infty$; the same is true
for $\partial \mathcal{R}{}_{-} / \partial r$ unless $\partial g_{tt} / 
\partial r$ goes to zero.  Correspondingly, if $\mathcal{S}$ is approached 
from $\mathcal{U}{}_{\mathrm{erg}}$, the potential $\mathcal{R}{}_-$ 
goes to $+ \infty$; the same is true for 
$\partial \mathcal{R}{}_{-} / \partial r$ unless $\partial g_{tt} / 
\partial r$ goes to zero. The potential $\mathcal{R}{}_+$ and 
its derivative are finite and continuous if $\mathcal{S}$ is crossed.

The statements of the previous paragraph are also true in the case 
that  $g_{t \varphi} > 0$ on $\mathcal{S}$, just with the potentials
$\mathcal{R}{}_+$ and $\mathcal{R}{}_-$ interchanged.

A further decomposition of $\mathcal{U}$ is necessary if $g_{t \varphi}$
takes positive and negative values on $\mathcal{S}$, which is possible
only if $\mathcal{S}$ is disconnected. We will not work out such cases 
here.

\subsection{Coordinate transformations}

We are free to make coordinate transformations of the form
\begin{equation}
t  \mapsto \tilde{t} = t + h(x^2,x^3) 
\, ,  \quad
\varphi \mapsto \tilde{\varphi} = \pm \varphi
\, , \quad
x^i \mapsto 
\tilde{x}{}^i = f^i  (x^2,x^3)
\label{eq:coord}
\end{equation}
where $h$, $f^2$ and $f^3$ are functions of $(x^2,x^3)$
that are arbitrary except for the condition that they define
an allowed coordinate transformation, i.e, that 
the determinant of the $(2 \times 2)$-matrix 
$(\partial f^i / \partial x^j)$ is non-zero.
  
Under such a transformaton, 
\begin{equation}
d\tilde{t} = dt +  
\dfrac{\partial h(x^2,x^3)}{\partial x^j} dx^j  
\, , \quad
d\tilde{\varphi} = \pm d\varphi 
\, , \quad
d\tilde{x}{}^i = \dfrac{\partial f^i(x^2,x^3)}{\partial x^j} dx^j  
\, ,
\end{equation}
hence
\begin{equation}
\dfrac{\partial}{\partial \tilde{t}} = 
\dfrac{\partial}{\partial t} 
\, , \quad
\dfrac{\partial }{\partial \tilde{\varphi}} =
\pm 
\dfrac{\partial}{\partial \varphi} 
\, , \quad
\dfrac{\partial f^j(x^2,x^3)}{\partial x^i} 
\dfrac{\partial}{\partial \tilde{x}{}^j } = 
\dfrac{\partial}{\partial x^i}
-
\dfrac{\partial h (x^2,x^3)}{\partial x^i} 
\dfrac{\partial }{\partial t} 
\, .  
\end{equation}
This implies that 
\begin{equation}
g_{tt} \mapsto \tilde{g}{}_{tt} = 
g_{tt} 
\, , \quad
g_{t \varphi} \mapsto \tilde{g}{}_{t \varphi} =
\pm \, g_{t \varphi} 
\, , \quad
g_{\varphi \varphi} \mapsto \tilde{g}{}_{\varphi \varphi} =
g_{\varphi  \varphi} 
\end{equation}
and, thus, 
\begin{equation}
g_{t \varphi}^2- g_{tt} g_{\varphi \varphi} \mapsto
\tilde{g}{}_{t \varphi} ^2 - \tilde{g}{}_{tt} \, \tilde{g}{}_{\varphi \varphi}
=g_{t \varphi}^2- g_{tt} g_{\varphi \varphi} \, .
\end{equation}
As also 
\begin{equation}
p_t \mapsto \tilde{p}{}_t = 
p_t
\, ,
\end{equation}
the potentials $\mathcal{R}{}_+$ and $\mathcal{R}{}_-$ remain
unchanged. Of course, after the transformation we have
to express them in terms of the new coordinates
$\big( f^2(x^2,x^3)), f^3(x^2,x^3) \big)$.

The transformations (\ref{eq:coord}) allow, e.g., 
in the Schwarzschild spacetime to switch from standard 
Schwarzschild coordinates to Eddington-Finkelstein or
Painlev{\'e}-Gullstrand coordinates. 

Note that, apart from a trivial rescaling of the time coordinate,
$t \mapsto k \, t$ with a constant $k$, the transformations 
(\ref{eq:coord}) are indeed the most general ones that
preserve all our assumptions. A transformation
of the form $t \mapsto t + \Omega_0 ^{-1} \varphi$ with
a non-zero constant $\Omega _0$ would leave each of the 
two potentials $\mathcal{R}{}_+$ and $\mathcal{R}{}_-$ 
invariant up to an additive constant, but it would violate the 
periodicity condition (\ref{eq:period}); the $\varphi$-lines 
would no longer be closed.

\subsection{Sign conventions}\label{subsec:sign}

There are several conventions that go into our definition of 
$\mathcal{R}{}_{\pm}$. Firstly, it is not necessary, though
convenient, to require that the force points into the direction
of the \emph{negative} gradient of the potential. This means 
that one could put a minus sign in front of one potential, or both.

Secondly, it is arbitrary which potential we label with a plus sign and 
which with a minus sign. We have chosen the signs such that 
$\dot{\varphi}$ is positive for the upper sign and negative for the
lower sign, see (\ref{eq:dotphi}). This, however, does not in
general determine the sign of the angular velocity 
\begin{equation}
\Omega = \dfrac{\dot{\varphi}}{\dot{t}}
\label{eq:Omega}
\end{equation}
because $\dot{t}$, which is given according to (\ref{eq:dott})
by

\begin{equation}
\dot{t} = - \dfrac{\omega _0}{g_{tt}} \mp 
\dfrac{g_{t \varphi}}{g_{tt}} 
\dfrac{
\sqrt{\omega _0 ^2 + \omega _p^2 g_{tt}}
}{
\sqrt{g_{t \varphi}^2 - g_{tt} g_{\varphi \varphi}}
}
\, ,
\label{eq:dott2}
\end{equation}
may have either sign. This is related to the question of whether
the chosen parametrisation is future-oriented or past-oriented
with respect to a time orientation. 

If $g_{\varphi \varphi} > 0$ (i.e., if there is no causality violation), 
the hypersurfaces $t = \mathrm{constant}$ are spacelike, i.e., 
$t$ is a time function which defines a time orientation. Then, if
$g_{tt} < 0$ (i.e., outside of an ergoregion), the first
term in (\ref{eq:dott2}) dominates the second one, so 
our convention $\omega _0 \ge 0$ implies that $\dot{t}$
is positive for both signs, so all light rays are future-oriented
and $\Omega$ has the same sign as $\dot{\varphi}$. 
(Recall that, actually,  the case $\omega _0 = 0$ is forbidden if
$g_{tt} < 0$.)   If, on the other hand, $g_{tt} > 0$ (i.e.,
inside an ergoregion), (\ref{eq:det2}) requires $g_{t \varphi}
\neq 0$, so the second term in (\ref{eq:dott2}) dominates
the first one. Our convention $\omega _0 \ge 0$ 
implies that $\dot{t} > 0$ if $g_{t \varphi} < 0$ and 
$\dot{t} < 0$ if $g_{t \varphi} > 0$. (Here the case 
$\omega _0 = 0$ is possible provided that $\omega _p
\neq 0$.)  So we have to keep in mind that inside an 
ergoregion our light rays are parametrised in a future-oriented 
way for one sign and in a past-oriented way for the other.  

If $g_{\varphi \varphi} < 0$ (i.e. in a causality-violating region), 
it depends on $\omega _p$ which term in (\ref{eq:dott2})
dominates, so no general statements are possible about the
sign of the angular velocity $\Omega$.

Thirdly, we could use, instead
of $\mathcal{R}{}_{\pm}$, a function of $\mathcal{R}{}_{\pm}$
that has the same critical points, e.g. the negative inverse whose
gradient is given by
\begin{equation}
\dfrac{\partial}{\partial x^i} 
\Bigg( \dfrac{-1}{\mathcal{R}{}_{\pm}} \Bigg)
=
\dfrac{1}{\mathcal{R}_{\pm}^2} \, \dfrac{\partial \mathcal{R}{}_{\pm}}{\partial x^i}
\, .
\label{eq:inverse}
\end{equation}
The relation between the potentials and their inverses can be read
from the identity
\begin{equation}
g_{tt} \, \mathcal{R}{}_+ \mathcal{R}{}_- =
\big( g_{t \varphi}^2 - g_{tt} g_{\varphi \varphi} \big)
\,  \omega _p ^2  
- g_{\varphi \varphi } \, \omega _0^2
\, .
\label{eq:R+R-}
\end{equation}
Note, however, that in our equation (\ref{eq:force}) the prefactor 
of the gradient on the right-hand side is strictly negative and 
finite at all points where (\ref{eq:det2}) and (\ref{eq:bound}) hold,
i.e., at all points where circular light rays may exist. This is no
longer true if we replace $\mathcal{R}{}_{\pm}$ by 
$-1/\mathcal{R}{}_{\pm}$: The new prefactor is still non-negative, 
but it is infinite at points where $\mathcal{R}{}_{\pm}$ has a zero.
This unwanted feature is avoided if we use $\mathcal{R}{}_{\pm}$
rather than its negative inverse. We also mention that in the vacuum
case, but not in a plasma, the inverse potentials give directly the 
angular velocity (\ref{eq:Omega}),
\begin{equation}
\Omega = 
 \dfrac{
 g_{tt} \, 
\sqrt{\omega _0^2+ \dfrac{\omega _p^2 g_{tt}}{\omega _0^2 c^2}}
}{
\mp \, \omega _0 \, 
\sqrt{g_{t \varphi} ^2  - g_{\varphi \varphi} g_{tt} } 
\, - \, g_{t \varphi} \,
\sqrt{\omega _0^2+ \omega _p^2 g_{tt}}
}
\, \underset{\omega _p \to 0}{\longrightarrow} \, 
\dfrac{\omega _0}{\mp \, \mathcal{R}{}_{\pm}}
\label{eq:Omega2}
\end{equation}

In the case $\omega _p=0$, the inverse potentials 
$1 / \mathcal{R}{}_{\pm}$ have been used for studying 
gravitational lensing in the Kerr-Newman spacetime by
Hasse and Perlick \cite{HassePerlick2006}. 
Using a different sign convention, Cunha et al 
\cite{CunhaBertiHerdeiro2017,CunhaHerdeiro2020,CunhaHerdeiroNovo2024} 
have utilised these potentials in their work 
on the existence of circular vacuum light rays in axially 
symmetric and stationary spacetimes. We will generalise 
their results to the plasma case in the next section.
The above arguments demonstrate that, in particular in
the presence of a plasma, it is more convenient and 
more natural to choose $\mathcal{R}{}_{\pm}$ rather
than their inverses.

\subsection{Properties of the potentials 
$\mathcal{R}{}_{\pm}$ near a horizon} 
\label{subsec:hor}

For analysing the properties of the potentials when a horizon 
is approached, we choose spherical polar coordinates 
$(t, \varphi , r , \vartheta)$ such that the horizon is 
represented as the hypersurface $r = r_h$. Then the
metric coefficients $g_{AB}$ and $g^{ij}$ and their 
derivatives are finite on the horizon.  As we are free
to change to horizon-penetrating coordinates, by
a transformation (\ref{eq:coord}), we may even
assume that \emph{all} metric coefficients $g_{\mu \nu}$
and $g^{\mu \nu}$ and their derivatives are finite on 
the horizon. 

A horizon is a lightlike hypersurface with finite 
circumference, so we must have 
\begin{equation}
g_{t \varphi} ^2 - 
g_{tt}  
g_{\varphi \varphi}  
= 0  
\, , \quad
g_{\varphi \varphi } > 0
\label{eq:horlim}
\end{equation}
at all points on the horizon with $0 < \vartheta < \pi$. 
We denote the domain of outer communication of the black hole
by $\mathcal{U}$. By definition, this is the open set adjacent
to the horizon where (\ref{eq:det2}) holds. We choose the
coordinate $r$ such that $r > r_h$ on $\mathcal{U}$.

In order to derive the limit behaviour of the potentials and their
derivatives at the horizon, we have to make sure that both
potentials are defined on a subset of $\mathcal{U}$ which
is adjacent to the horizon. This is achieved by requiring that
$- \omega _p ^2 \, g_{tt} $ is bounded on a neighbourhood
of the horizon and that we have chosen a frequency constant 
$\omega _0$ such that $\omega _0^2 >
\mathrm{sup} \big(- \omega _p^2 \,  g_{tt} \big)$.
In addition we also have to assume that the condition
\begin{equation}
\sqrt{g_{t \varphi} ^2 - g_{tt} g_{\varphi \varphi}}
\, 
\dfrac{\partial \omega _p^2}{\partial r}
 \to 0
\label{eq:horgrad}
\end{equation}
holds  for $r \to r_h$. This is certainly true if the gradient of 
$\omega _p^2$ is bounded near the horizon. 

For deriving the limit of $\mathcal{R}{}_{\pm}$ for $r \to r_h$ we 
first observe that
\begin{equation}
\dfrac{1}{\mathcal{R}{}_{\pm}} =
\dfrac{
\mathcal{R}{}_{\mp}
}{
\mathcal{R}{}_{\pm} \mathcal{R}{}_{\mp}
}
=
\dfrac{
\mp g_{t \varphi} \omega _0 +
\sqrt{g_{t \varphi}^2 - g_{tt} g_{\varphi \varphi}}
\sqrt{\omega _0 ^2 + \omega _p ^2 g_{tt}}
}{
-\omega _0^2 g_{\varphi \varphi}
+
\omega _p ^2 
\big(g_{t \varphi}^2 - g_{tt} g_{\varphi \varphi} \big)
}
\, .
\label{eq:hor0}
\end{equation}
If we exclude the case $\omega _0 = 0$, which will be 
considered at the end of this section, we find with the help
of our assumption that $\omega _0^2 > \mathrm{sup}
\big( - \omega _p ^2 \, g_{tt} \big)$ that
\begin{equation}
\dfrac{1}{\mathcal{R}{}_{\pm}} \mp
\dfrac{g_{t \varphi}}{g_{\varphi \varphi} \,  \omega _0 }
\rightarrow - 0
\label{eq:hor1}
\end{equation}
for $r \to r_h$. Here we write $-0$ to indicate that the limit is 
approached from below. 

To evaluate this result further, we first consider the case of a 
non-rotating horizon,  i.e., we assume that $g_{t \varphi} = 0$ 
and thus $g_{tt}=0$ on the horizon. From 
(\ref{eq:hor1}) we read that then
\begin{equation}
\mathcal{R}{}_{\pm} \rightarrow - \infty
\label{eq:horlim1}
\end{equation}
for $r \to r_h$. This 
case covers all spherically symmetric and static black holes and 
also, e.g.,  
the NUT metric. In the latter case $g_{t \varphi}$ is non-zero in 
general, but it is zero on the horizon. 

The situation is different if the horizon rotates, i.e., if $g_{t \varphi}
\neq 0$ and hence $g_{tt} >0$ on the horizon.  From 
(\ref{eq:hor1}) we read that then 
\begin{equation}
\mathcal{R}{}_{\pm} \rightarrow 
\pm \dfrac{g_{t \varphi}}{g_{\varphi \varphi} \omega _0} \Bigg| _{r= r_h}
\, ,
\label{eq:horlim2}
\end{equation} 
i.e., the potentials approach finite and non-zero values that are 
equal in magnitude and opposite in sign. This case covers all 
metrics that describe stationarily rotating black holes such as the 
Kerr metric.

Knowing the limit behaviour of the potentials, we now derive the
limit behaviour of their derivatives. We will see that in this case  
it makes a difference whether the horizon is degenerate
or non-degenerate. We start out from the equations

\begin{equation}
\dfrac{1}{\mathcal{R}{}_-} + \dfrac{1}{\mathcal{R}{}_+}
=
\dfrac{\mathcal{R}{}_+ +\mathcal{R}{}_-}{
\mathcal{R}{}_+\mathcal{R}{}_-}
=
\dfrac{
2 \, \sqrt{g_{t \varphi}^2 - g_{tt} g_{\varphi \varphi}}
\sqrt{\omega _0 ^2 + \omega _p^2 g_{tt}}
}{ 
- g_{\varphi \varphi} \omega _0^2 + \omega _p ^2
\big( g_{t \varphi}^2 - g_{tt} g_{\varphi \varphi} \big)
}
\end{equation}
and

\begin{equation}
\dfrac{1}{\mathcal{R}{}_-} - \dfrac{1}{\mathcal{R}{}_+}
=
\dfrac{\mathcal{R}{}_+ - \mathcal{R}{}_-}{
\mathcal{R}{}_+\mathcal{R}{}_-}
=
\dfrac{
\pm 2 \, g_{t \varphi} \omega _0
}{ 
- g_{\varphi \varphi} \omega _0^2 + \omega _p ^2
\big( g_{t \varphi}^2 - g_{tt} g_{\varphi \varphi} \big)
}
\end{equation}
which follow directly from (\ref{eq:R}). 
Differentiating these equations, and excluding 
again the case $\omega _0 = 0$,  shows that
\begin{equation}
\dfrac{1}{\mathcal{R}{}_+^2}
\dfrac{\partial \mathcal{R}{}_+}{\partial r}
+
\dfrac{1}{\mathcal{R}{}_-^2}
\dfrac{\partial \mathcal{R}{}_-}{\partial r}
+
\dfrac{2 \,
\sqrt{\omega _0 ^2 + \omega _p ^2 g_{tt}}
}{
g_{\varphi \varphi} \omega _0^2 
}
\, 
\dfrac{\partial 
\sqrt{ g_{t \varphi}^2 - g_{tt} g_{\varphi \varphi}}
}{
\partial r
}
\rightarrow 0
\label{eq:inv1}
\end{equation}
and

\begin{equation}
\dfrac{1}{\mathcal{R}{}_+^2}
\dfrac{\partial \mathcal{R}{}_+}{\partial r}
-
\dfrac{1}{\mathcal{R}{}_-^2}
\dfrac{\partial \mathcal{R}{}_-}{\partial r}
\quad 
\text{stays finite}
\label{eq:inv2}
\end{equation}
if the horizon is approached. Here we have used  our
assumptions that $\omega _0 ^2 > \mathrm{sup} 
\big( - \omega _p ^2 \, g_{tt}\big)$ and that
(\ref{eq:horgrad}) holds.

If the horizon is non-degenerate, 
$\partial \sqrt{g_{t \varphi}^2 - g_{tt} g_{\varphi \varphi}}
/ \partial r$ goes to $+\infty$ for $r \to r_h$. (Recall
that we approach the horizon from the side where $r > r_h$.)
As $\mathcal{R}{}_{\pm}$ does not go to 0, (\ref{eq:inv1})
and (\ref{eq:inv2}) imply that

\begin{equation}
\dfrac{\partial \mathcal{R}{}_{\pm}}{\partial r}
\rightarrow - \infty
\label{eq:hor3}
\end{equation}
for $r \to r_h$.

If the horizon is degenerate, 
$\partial \sqrt{g_{t \varphi}^2 - g_{tt} g_{\varphi \varphi}}
/ \partial r$ approaches a finite and positive value  for 
$r \to r_h$. In the non-rotating case, where
$\mathcal{R}{}_{\pm}^2  \to \infty$, (\ref{eq:inv1}) and 
(\ref{eq:inv2}) imply that again (\ref{eq:hor3}) holds. In
the rotating case, however, where (\ref{eq:horlim2})
holds, (\ref{eq:inv1}) and (\ref{eq:inv2}) imply that 
the derivatives of the potentials approach finite values.

Finally, we consider the case $\omega _0 = 0$ which was 
left out so far. In this case the potentials, which are defined near
the horizon only if $g_{tt} > 0$ and $\omega _p^2 > 0$, 
reduce to
\begin{equation}
\mathcal{R}{}_+ = \mathcal{R}{}_- =
\dfrac{\omega _p}{\sqrt{g_{tt}}} \, 
\sqrt{g_{t \varphi } ^2 - g_{tt} g_{ \varphi \varphi}}
\, .
\end{equation}
From this expression and from its $r$-derivative we read that 
no general statements about the limit behaviour of the  
potentials and their derivatives are possible: Apart from the
obviuous fact  that the limit of $\mathcal{R}{}_+ =
\mathcal{R}{}_-$ cannot be negative, everything is possible,
depending on the limit behaviour of $g_{tt}$,
$\omega _p$ and $\partial \omega _p/ \partial r$. 
 
So we see that the case that $\omega _0 = 0$ and the
case that the horizon is rotating and degenerate are special. 
In all other cases $\partial \mathcal{R}{}_{\pm} / \partial r$ 
goes to  $- \infty$ if a horizon is approached. This implies
that there is a positive $r_0 (\vartheta )$ such that neither
the potential $\mathcal{R}{}_{\pm}$ itself nor any small
perturbation thereof has a critical point in the domain
\begin{equation}
\mathcal{U}{}_{r_0} =
\big\{
(r \, \mathrm{sin} \, \vartheta , r \, \mathrm{cos} \, \vartheta )
\, \big| \, r_h < r < r_h + \delta (\vartheta ) \, , \, 0 < \vartheta< \pi \, \big\}
\, ,
\end{equation}
or on its boundary. Here the case that 
$\delta ( \vartheta ) \to 0$ for $\mathrm{sin} \, \vartheta \to 0$ 
cannot be excluded. We keep in mind that this is 
true not only for vacuum light rays but also in a plasma, 
provided that $\omega _0 ^2 > 
\mathrm{sup}\big(-  \omega _p ^2 \, g_{tt} \big)$  
and that (\ref{eq:horgrad}) holds.

\subsection{Properties of the potentials $\mathcal{R}{}_{\pm}$ 
on an axis}\label{subsec:ax}

In order to analyse the properties of the potentials near a 
regular axis of symmetry, we use cylindrical polar 
coordinates $(t, \varphi , \rho , z)$ such that the axis
is represented as the boundary $\rho = 0$ of the 
half-plane $\rho > 0$. The axis is called regular if
the metric satisfies the condition of ``elementary 
flatness'' there. This means that the coordinates
can be chosen such that on the interval 
$z_1<z<z_2$ where the axis is regular there is 
a strictly positive function $f(z)$ such that
\begin{equation}
g_{tt} \rightarrow -f(z)^2 
\, , \quad
\dfrac{g_{t\varphi}}{\rho} \rightarrow  0 
\, , \quad
\dfrac{g_{ \varphi \varphi}}{\rho ^2} \rightarrow  1 \, , 
\label{eq:limax1}
\end{equation}
\begin{equation}
\dfrac{\partial g_{tt}}{\partial \rho} \rightarrow 0 
\, , \quad
\, \dfrac{\partial g_{t \varphi}}{\partial \rho} 
\rightarrow 0 
\, , \quad
\dfrac{1}{\rho} 
\dfrac{\partial g_{\varphi \varphi}}{\partial \rho} \rightarrow 2 
\, ,
\label{eq:limax2}
\end{equation}
for $\rho \to 0$. This implies that
\begin{equation}
\dfrac{1}{\rho } \, \sqrt{g_{t \varphi} ^2 - g_{tt} g_{\varphi \varphi}}
\to f(z)\, , \quad
\dfrac{ 
\partial \sqrt{g_{t \varphi} ^2 - g_{tt} g_{\varphi \varphi}}
}{
\partial \rho}
\to f(z)  \, , \quad
\end{equation}
for $\rho \to 0$. 
   
We have to make sure that the domain
of definition of $\mathcal{R}{}_{\pm}$ extends to the axis.
This requires that we have to assume that 
$ - \omega _p ^2 \, g_{tt}$ is bounded near the axis, 
and we have to choose a frequency constant $\omega _0$
such that $\omega _0^2 > 
\mathrm{sup} \big( - \omega _p ^2 \, g_{tt}\big)$. 
Moreover, we have to require that  
\begin{equation}
\rho \, \dfrac{\partial \omega _p^2}{\partial \rho} \to 0
\label{eq:axgrad}
\end{equation}
for $\rho \to 0$. This condition, which is analogous to 
(\ref{eq:horgrad}), is certainly true if the gradient of 
$\omega _p^2$ is bounded near the axis. 
 

Then the axis is on the boundary of the domain 
where the potential $\mathcal{R}{}_{\pm}$ 
is defined, for both signs, and we find
\begin{equation}
\mathcal{R}{}_{\pm} 
\rightarrow 0 \, , \quad
\dfrac{\partial \mathcal{R}{}_{\pm}}{\partial \rho}
+ \dfrac{1}{f(z)} \,\sqrt{\omega _0^2 + \omega _p ^2 g_{tt}} 
\rightarrow 0
\label{eq:axR}
\end{equation}
for $\rho \to 0$. This is true for all values of $z$ in the interval 
$z_1<z<z_2$ where the axis is regular. 

This observation has important consequences for 
spacetimes that are asymptotically flat, see Section \ref{subsec:inf} 
below. If this is true, and if the axis is regular on an interval
$z_1 < z < \infty$, we have $f(z) \to 1$ for $z \to \infty$.
As the square-root in (\ref{eq:axR}) is strictly positive, the gradient of 
$\mathcal{R}{}_{\pm}$ is bounded away from zero near 
the axis, i.e., there is a constant $\rho _0$ such
that neither the potential $\mathcal{R}{}_{\pm}$ itself nor any 
small perturbation thereof has a critical point in the domain
$0 < \rho < 2 \, \rho _0$, $z_1 < z < \infty$. Similarly, if the axis
is regular on the interval $- \infty < z < z_2$, asymptotic flatness
implies that $f(z) \to 1$ for $z \to - \infty$, so there is a constant 
$\rho _0$ such that neither the potential itself nor any small 
perturbation thereof can have a critical point on the domain  
$0 < \rho < 2 \, \rho _0$, $- \infty < z < z_2$. 
We will later apply this result to asymptotically flat spacetimes 
where the entire axis is regular, i.e. $z_1 = - \infty$ and 
$z_2 = \infty$, and to asymptotically flat black-hole
spacetimes with horizon at $\sqrt{\rho ^2 + z^2} = r_h$
where the axis is regular between the horizon
and infinity, i.e. the result is true for any $z _1 \ge r_h$ and 
any $z_2 \le - r_h$. Note that the limiting values $z_1 =
r_h$ and $z _2 = - r_h$ are indeed included because $f(z)$
does not go to $\infty$ for $z \to \pm r_h$: It goes to 0 for a 
non-rotating horizon and to a non-zero finite value for a 
rotating horizon.

\subsection{Properties of the potentials $\mathcal{R}{}_{\pm}$ 
at infinity}\label{subsec:inf}

We now turn to an investigation of the behaviour of
the potentials at infinity. We assume that the spacetime
is asymptotically flat in the sense that we may choose
spherical polar coordinates $(t , \varphi , r , \vartheta )$ 
such that
\begin{equation}
g_{tt} \rightarrow -1 
\, , \quad
\dfrac{g_{t\varphi}}{r \, \mathrm{sin} \, \vartheta} \rightarrow  0 
\, , \quad
\dfrac{g_{ \varphi \varphi}}{r ^2 \, \mathrm{sin} ^2 \vartheta} \rightarrow  1 
\, , 
\label{eq:liminf1}
\end{equation}
\begin{equation}
r \, \dfrac{\partial g_{tt}}{\partial r} \rightarrow 0 
\, , \quad
\dfrac{\partial g_{t \varphi}}{\partial r} \rightarrow 0 
\, , \quad
\dfrac{1}{r \, \mathrm{sin} ^2 \vartheta } 
\dfrac{\partial g_{\varphi \varphi}}{\partial r} \rightarrow 2 
\label{eq:liminf2}
\end{equation}
for $r \to \infty$ and $0 < \vartheta < \pi$. This implies that
\begin{equation}
\dfrac{
\sqrt{g_{t \varphi} ^2 - g_{tt} g_{\varphi \varphi}}
}{
r \, \mathrm{sin} \vartheta
}
\to 1 \, , 
\quad 
\dfrac{ 
\partial \sqrt{g_{t \varphi} ^2 - g_{tt} g_{\varphi \varphi}}
}{
\partial r}
\to \mathrm{sin} \, \vartheta \, .
\end{equation}
We have to assume  
that $- \omega_p ^2 \, g_{tt} $ is bounded at infinity, and we 
have to choose a frequency constant $\omega _0$ with 
$\omega _0 ^2 >\mathrm{sup}\big(- \omega _p ^2 \, g_{tt} \big)$. 
Moreover, 
we have to require that
\begin{equation}
\sqrt{g_{t \varphi}^2 - g_{tt} g_{\varphi \varphi}}
\, \dfrac{\partial \omega _p^2}{\partial r} \to 0
\label{eq:infgrad}
\end{equation}
for $r \to \infty$ which is certainly true if the gradient 
of $\omega _p^2$ falls off stronger than $1/r$. 

Then it is straight-forward to verify that
\begin{equation}
\mathcal{R}{}_{\pm} ( r , \vartheta)
\rightarrow - \infty \, , \quad
\dfrac{\partial \mathcal{R}{}_{\pm} ( r , \vartheta)}{\partial r}
+ \mathrm{sin} \, \vartheta \, 
\sqrt{\omega _0^2 +\omega _p( r , \vartheta )^2 g_{tt} ( r , \vartheta )} 
\rightarrow 
 0 
 \label{eq:infR}
\end{equation}
for $ r \to \infty$. As the square-root is bounded away from zero 
this guarantees that there is an $r_0 ( \vartheta )$ such that neither 
the potential $\mathcal{R}{}_{\pm}$ itself nor any small 
perturbation thereof has a critical point in the domain
$r_0 (\vartheta )/2  < r < \infty$, $0 < \vartheta < \pi$. We
cannot exclude the possibility that $r_0 ( \vartheta ) \to \infty$
for $\mathrm{sin} \, \vartheta \to 0$.

\section{Existence theorems for circular light rays}
\label{sec:theorems}

Cunha et al. 
\cite{CunhaBertiHerdeiro2017,CunhaHerdeiro2020,CunhaHerdeiroNovo2024} 
have proven existence theorems for circular
light rays in vacuum (i.e. for circular lightlike geodesics) in
axially symmetric and stationary spacetimes. To that end they 
have used two potentials $H_{\pm}$ that are related to
our potentials $\mathcal{R}{}_{\pm}$ by the equation
$H_{\pm} = \mp 1/ \mathcal{R}{}_{\pm}$ if the latter 
are restricted to the case that $\omega _p =0$. The
reasons why we use $\mathcal{R}{}_{\pm}$, rather
than $H_{\pm}$, have been outlined in Section
\ref{subsec:sign}. The fact that $\mathcal{R}{}_{\pm}$
is well defined also for $\omega _p \neq 0$ allows us
to generalize the results of Cunha et al. to the plasma
case. 

Following Cunha et al. closely, we utilise the Brouwer degree, 
which is a standard concept in differential topology, and the 
fact that it is a homotopy invariant. The Brouwer degree 
is defined for a smooth ($C^2$ would do) function 
$f$ from a compact manifold $\mathcal{K}$ to a manifold
$\mathcal{N}$ of the same dimension. In our case 
$f$ is the gradient of the potential $\mathcal{R}{}_{\pm}$, 
$\mathcal{K}$ is the closure of an open  and bounded
subset of the connected domain $\mathcal{V}{}_{\pm}$ 
where the potential $\mathcal{R}{}_{\pm}$ is defined, 
and $\mathcal{N}$ is $\mathbb{R}{}^2$. One has
to assume that 0 is a regular value of $f$. In our case,
this means that $\mathcal{R}{}_{\pm}$ is a Morse function, 
i.e. that at all critical points of the potential its Hessian is 
non-degenerate. As we have the positive definite metric
$g^{ij}$ at our disposal, this assumption is
tantamount to the condition that the $(2 \times 2)$-matrix 
$\big( g^{ik} \partial^2 \mathcal{R}{}_{\pm} / 
\partial x^k \partial x^j \big)$ has two non-zero
eigenvalues, so its determinant is non-zero. Following
the terminology of Cunha et al., we assign to each
critical point a ``topological charge'' which, by 
definition, is the sign of this determinant. In other 
words, the topological charge is $+1$ if the
critical point is an extremum (i.e., a local minimum
or a local maximum), while it is $-1$ if the critical
point is a saddle. The Brouwer degree of the map 
$f$ is defined as the sum of the topological charges
over all critical points. If the potential has no critical points,
it is a Morse function and one assigns the Brouwer degree 
0 to its gradient. 
Now the homotopy invariance of the degree says
that two maps $f^0: \mathcal{K} \rightarrow \mathcal{N}$ 
and $f^1: \mathcal{K} \rightarrow \mathcal{N}$ have 
the same degree if they can be deformed into each other
by a one-parameter family of maps
$f^{\varepsilon} : \mathcal{K} \rightarrow \mathcal{N}$,
continuously depending on $\varepsilon \in [0,1]$,
such that $f^{\varepsilon}$ does not take the value 0
on the boundary of $\mathcal{K}$, for all $\varepsilon 
\in [0,1]$. While every textbook on 
differential topology discusses the homotopy invariance
of the Brouwer degree, some of them restrict to the case that 
$\mathcal{K}$ is closed, i.e., compact without boundary. 
A complete proof for the case that $\mathcal{K}$ is
compact with boundary, which is the case of relevance
to us, can be found e.g. in the book by Dinca and Mawhin 
\cite{DincaMawhin2021}.

If applied to our potentials $\mathcal{R}_+$ and 
$\mathcal{R}{}_-$, the homotopy invariance of
the Brouwer degree gives us the following result.

\begin{proposition}
Consider a one-parameter family of axially symmetric and
stationary spacetimes with metrics $g_{\mu \nu}^{\varepsilon}$ 
and axially symmetric and stationary plasma frequencies
$\omega _p ^{\varepsilon}$, depending continuously on 
a parameter $\varepsilon \in [0,1]$. Let 
$\mathcal{R}_{\pm}^{\varepsilon}$ be the corresponding 
potentials, defined on the open and connected subsets 
$\mathcal{V}_{\pm}^{\varepsilon}$. Assume that
there is a family of compact subsets $\mathcal{K}_{\pm}^{\varepsilon}$
and a family of diffeomorphisms $\mathcal{K}_{\pm}^{\varepsilon}
\rightarrow \mathcal{K}_{\pm}^0$, depending continuously on
$\varepsilon \in [0,1]$, such that all critical points of
$\mathcal{R}{}_{\pm} ^{\varepsilon}$ 
are contained in the interior of 
$\mathcal{K_{\pm}^\varepsilon}$. If $\mathcal{R}_{\pm}^0$ 
and $\mathcal{R}{}_{\pm}^1$ are Morse functions, then 
the number of critical points of $\mathcal{R}_{\pm}^1$
differs from the number of critical points of       
$\mathcal{R}_{\pm}^0$ by an even number $2n$, where
$n$ of the additional critical points are saddles and the other 
$n$ are extrema.
\label{prop:Brouwer}
\end{proposition}

When applying this proposition one has to make sure that 
$\mathcal{R}{}_{\pm} ^{\varepsilon}$ satisfies all the 
assumptions. The proposition is not applicable if  
$\mathcal{R}_{\pm}^0$ or $\mathcal{R}{}_{\pm}^1$
fails to be a Morse function. Then its critical points need
not be isolated, i.e., there may be a continuum of critical
points which means that the number of critical points is
infinite and not even countable. Morse functions are generic
in the sense that every non-Morse function becomes a 
Morse function under an appropriate small perturbation.
Also, the proposition is not applicable if both functions 
$\mathcal{R}_{\pm}^0$  and $\mathcal{R}{}_{\pm}^1$ 
are Morse functions but if one of them has infinitely many 
critical points. As critical points are isolated if they 
are non-degenerate, it is not possible that infinitely many
of them are contained in a compact set. Finally, it is possible 
that $\mathcal{R}{}_{\pm} ^1$ and $\mathcal{R}{}_{\pm}^0$
are Morse functions all of whose critical points lie in the interior 
of a compact set $\mathcal{K}_{\pm}^1$, but that 
it is impossible to find a deformation
$\mathcal{R}{}_{\pm}^{\varepsilon}$
and compact sets $\mathcal{K}{}_{\pm}^{\varepsilon}$,
both depending continuously on $\varepsilon$, such 
that the critical points of $\mathcal{R}{}_{\pm}^{\varepsilon}$
are in the interior of $\mathcal{K}{}_{\pm}^{\varepsilon}$
for all $\varepsilon \in [0,1]$; this situation happens if,
intuitively speaking, $\mathcal{R}{}_{\pm}^1$
has a critical point at infinity. Examples for all three cases
will be given in Section \ref{subsec:Minkowski} below.  

In their first paper Cunha et al. \cite{CunhaBertiHerdeiro2017}
considered the case that $\omega _p =0$ throughout, 
and they applied what we have formulated as Proposition \ref{prop:Brouwer}
to the case that $g_{\mu \nu}^0$ is the metric of a spacetime without 
circular light rays and  $g_{\mu \nu}^1$ is the metric of a spacetime 
with circular light rays. The interpretation is that $g_{\mu \nu}^0$
describes the spacetime of a regular star before 
it undergoes gravitational collapse, while $g_{\mu \nu}^1$ describes
the corresponding spacetime after the star has collapsed to an 
ultracompact object that is not a black hole. The crucial 
observation is that, in our notation, circular light rays 
of the metric $g_{\mu \nu}^1$ come in pairs, always a saddle 
together with an extremum. In their follow-up 
papers, Cunha et al. \cite{CunhaHerdeiro2020,CunhaHerdeiroNovo2024}
investigated the Brouwer degree of the gradients of their 
potentials for black-hole spacetimes. 

Our generalisation to the plasma case allows for applications
where both the spacetime metric and the plasma density are
deformed. We will concentrate, however, in the following 
on one-parameter families of the form
\begin{equation}
\mathcal{R}{}_{\pm} ^{\varepsilon} =
\dfrac{g_{t \varphi}}{g_{tt}} \omega _0 
+ \dfrac{1}{g_{tt}} \sqrt{g_{t \varphi} ^2
- g_{tt} g_{\varphi \varphi}}
\sqrt{\omega _0 ^2 +
\varepsilon \omega _p ^2 g_{tt}}
\label{eq:Repsilon}
\end{equation}
where the spacetime metric is kept fixed. 
$\mathcal{R}{}_{\pm}^{\varepsilon}$ is defined on
the domain $\mathcal{V}{}_{\pm}$ where 
$\mathcal{R}{}_{\pm}^1$ is defined, for all 
$\varepsilon \in [ 0 , 1]$, if $-\omega _p ^2 \, g_{tt}$ 
is bounded on $\mathcal{V}{}_{\pm}$ and 
$\omega _0^2 > \mathrm{sup} \big(- \omega _p ^2 \, g_{tt} \big)$.
In all applications we will assume that this is the 
case. For other values of $\omega _0$ one could 
apply Proposition \ref{prop:Brouwer} as well, but 
one would have to consider domains 
$\mathcal{V}{}_{\pm} ^{\varepsilon}$ that depend
on $\varepsilon$ and possibly also frequency constants
$\omega _0$ that depend on $\varepsilon$. 

Moreover, we have to make sure that 
$\mathcal{R}{}_{\pm}^1$ and $\mathcal{R}_{\pm}^0$
are Morse functions and that all critical points of 
$\mathcal{R}{}_{\pm}^{\varepsilon}$ are contained
in the interior of a compact set $\mathcal{K}$.  
(If all $\mathcal{R}{}_{\pm}^{\varepsilon}$ are defined
on the same domain $\mathcal{V}{}_{\pm}$, there is 
no need to consider $\varepsilon$-dependent compact sets.) 
Then Proposition \ref{prop:Brouwer}
guarantees that the circular light rays in the plasma
density $\omega _p^2$ differ from the circular vacuum 
light rays on the same spacetime by $n$ saddles and
$n$ extrema. This implies, in particular, that there
is at least one circular light ray in the plasma if there
is an odd number of circular light rays in vacuum. In 
other words, by introducing a plasma we cannot destroy all 
circular light rays if their number is odd. 

Note that Proposition \ref{prop:Brouwer} has to be
applied to $\mathcal{R}{}_{+}$ and $\mathcal{R}{}_-$
separately and that these two potentials may have 
different domains of definition.

In the following sections we will consider several specific 
axially symmetric and stationary spacetimes.  We will 
discuss the general results for circular light rays that  
can be concluded from Proposition \ref{prop:Brouwer} 
and we will illustrate this with particular plasma 
densities. 

\section{Regular spacetimes}
\label{sec:reg}

In this section we exemplify Proposition \ref{prop:Brouwer}
with the case that the axially symmetric and stationary spacetime 
and the axially symmetric and stationary plasma density under
coonsideration are regular. By that we mean that the metric 
and the plasma density are defined on the entire half-plane 
\begin{equation}
\hspace{-1.75cm}
\mathcal{U} =
\{ (\rho , z ) 
\, | \, \rho >0 \, , \, - \infty < z < \infty \}
=
\{ ( r \, \mathrm{sin} \, \vartheta , r \, \mathrm{cos} \, \vartheta ,  ) 
\, | \, r >0 \, , \, 0 < \vartheta < \pi \}
\, ,
\label{eq:halfplane}
\end{equation} 
that (\ref{eq:det2}) holds on $\mathcal{U}$, 
that $\rho = 0$ is a regular axis (recall Section \ref{subsec:ax}), 
that the spacetime is asymptotically flat (recall Section
\ref{subsec:inf}), and that the plasma density is bounded
on the half-plane $\mathcal{U}$. 
These assumptions allow us to choose a frequency constant 
$\omega _0$ such that $\omega _0 ^2 > \mathrm{sup} \big(
- \omega _p ^2 \, g_{tt} \big)$. If we connect with 
the vacuum case on the same spacetime by defining the 
one-parameter family of potentials (\ref{eq:Repsilon}), we 
find that $\mathcal{R}{}_{\pm}^{\varepsilon}$ is defined
on the entire halfplane $\mathcal{U}$ for all $\varepsilon \in [ 0 , 1 ]$. 

If in addition the conditions (\ref{eq:axgrad}) and 
(\ref{eq:infgrad}) hold
for $r \, \mathrm{sin} \, \vartheta \to 0$ and $r \to \infty$, 
respectively, the results of Sections \ref{subsec:ax} 
and \ref{subsec:inf} guarantee that there is a compact 
subset $\mathcal{K}$ of $\mathcal{U}$ such that all 
critical points of $\mathcal{R}{}_{\pm}^{\varepsilon}$ 
lie in the interior of $\mathcal{K}$, for all $\varepsilon 
\in [0,1,]$. In the notation of Sections \ref{subsec:ax} 
and \ref{subsec:inf}, this compact set is of the form 
\begin{equation}
\mathcal{K} = 
\{ ( r \, \mathrm{sin} \, \vartheta , r \, \mathrm{cos} \, \vartheta ,  ) 
\, | \, r \, \mathrm{sin} \vartheta \ge \rho _0 \, , \, r \le r_0 ( \vartheta )
\}
\, .
\end{equation}
If $\mathcal{R}{}_{\pm}^0$ and $\mathcal{R}{}_{\pm}^1$ 
are Morse functions, which is generically true, Proposition
\ref{prop:Brouwer} guarantees that the 
circular light rays in the plasma density $\omega _p^2$
differ from the circular light rays in vacuum by $n$ saddles
and $n$ extrema. This result can be applied, e.g., to the 
spacetime of a star that is not ultracompact or to 
Minkowski spacetime. In both cases $\mathcal{R}{}_{\pm}^0$ 
has no critical points, so in the plasma the number of circular light
rays must be even. 

In the following we
illustrate this situation with plasma densities on Minkowski
spacetime. We choose three examples where
the assumptions of Proposition \ref{prop:Brouwer} are
\emph{not} satisfied, just to demonstrate in which situations 
this is the case.
 
\subsection{Minkowski spacetime}
\label{subsec:Minkowski}

In this section we give three examples for plasma densities on
Minkowski spacetime. Note that then the light rays in the 
plasma are timelike geodesics of the conformally rescaled
metric $\omega _p^2 g_{\mu \nu}$, i.e., they mimic the
motion of freely falling massive particles in this rescaled
metric. In this sense, light propagation in a plasma on
Minkowski spacetime may be viewed as an example of
\emph{analogue gravity}.

In each of the three following examples we choose a
frequency constant such that the potentials (\ref{eq:Repsilon}) 
are defined on the entire half-plane (\ref{eq:halfplane}), but 
that the assumptions of Proposition \ref{prop:Brouwer} are 
violated in three  different ways: In the first example 
$\mathcal{R}{}_{\pm} = \mathcal{R}{}_{\pm}^1$ fails to 
be a Morse function, in the second it is a Morse function but it has 
infinitely many critical points (which is possible because in this
example the gradient of the potential is unbounded)  
and in the third we choose $\omega _0^2$ equal to the 
supremum of $- \omega _p^2 \, g_{tt}$, rather than strictly 
bigger. 

In the first example we consider Minkowski spacetime
in cylindrical polar coordinates $(t, \varphi, \rho, z)$,
\begin{equation}
g_{tt} = -1 \, , \quad
g_{\varphi \varphi} = \rho ^2  
\, , \quad  
g_{\rho \rho} = 1 \, \quad
g_{zz} = 1 \, .
\label{eq:Minkowski}
\end{equation}
All other metric coefficients are zero.
Then the two potentials coincide, 
\begin{equation}
\mathcal{R}{}_+ ( \rho , z )
=
\mathcal{R}{}_- (\rho  , z )
= 
- \rho
\, \sqrt{\omega _0 ^2 - \omega _p( \rho, z )^2}
\, .
\end{equation}
We choose the plasma density
\begin{equation}
\omega _p ( \rho , z ) ^2= \omega _c ^2
\, \mathrm{sin} ^2 \dfrac{\rho}{\rho _0}
\end{equation}
where $\omega _c$ is a constant with the dimension of a 
frequency and $\rho _0$ is a constant with the dimension of 
a length. Then $- \omega _p ^2\, g_{tt} $ is bounded on
the half-plane $\mathcal{U}$, with $\mathrm{sup} 
\big( - \omega _p ^2\, g_{tt} \big) =\omega _c ^2$. 
We choose a frequency constant $\omega _0 > \omega _c$ 
which guarantees that 
\begin{equation}
\mathcal{R}{}_+ (\rho , z)
=
\mathcal{R}{}_- (\rho , z) 
= 
- \rho
\, \sqrt{\omega _0 ^2 - \omega _c ^2
\, \mathrm{sin} ^2 \dfrac{\rho}{\rho _0}} \, ,
\end{equation}
is defined on the entire half-plane $\mathcal{U}$. 
However, there is a continuum of critical points, located on 
infinitely many vertical lines in $\mathcal{U}$ which are 
given by the equation
\begin{equation}
\omega _c ^2  \, \mathrm{sin} \dfrac{\rho}{\rho _0}
\Big(  \mathrm{sin} \dfrac{\rho}{\rho _0}
+ 
\dfrac{\rho}{\rho _0} \, \mathrm{cos} \dfrac{\rho}{\rho _0} \, 
\Big) = \omega _0 ^2 \, ,
\end{equation}
so the potential fails to be a Morse function and 
Proposition \ref{prop:Brouwer} is not applicable.

\begin{figure}
\centerline{\includegraphics[width=0.6 \textwidth]{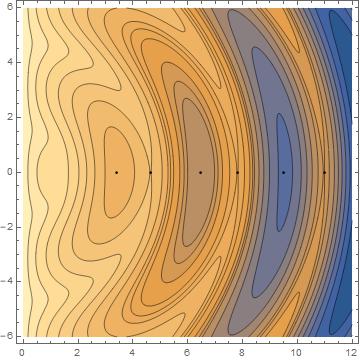}}
\caption{
This picture shows the equipotential lines 
$\mathcal{R}{}_+ = \mathcal{R}{}_- = \mathrm{constant}$ 
for the plasma density (\ref{eq:Minkomegap}) on Minkowski
spacetime for $\omega _0 = \sqrt{4/3} \,  \omega _c$ in the 
$( r \, \mathrm{sin} \, \vartheta , 
r \, \mathrm{cos} \, \vartheta )$-halfplane $\mathcal{U}$. $r_0$ is chosen 
as the unit on the axes. The circular light rays are marked by 
black dots. There are infinitely many saddles with local maxima in
between.   The saddles are minima (stable) in $r$ direction and
maxima (unstable) in $\vartheta$ direction. The potential goes,
in an oscillatory fashion, to $- \infty$ for 
$r \, \mathrm{sin} \, \vartheta \to \infty$. 
} 
\label{fig:Minkowski1}
\end{figure}

For the second (equally contrived) example we consider
again Minkowski spacetime,  this time in spherical
polar coordinates $(t , \varphi , r , \vartheta )$,
with the non-zero metric coefficients
\begin{equation}
g_{tt} = - 1 \, , \quad
g_{\varphi \varphi} = r^2 \mathrm{sin} ^2 \vartheta \, , \quad
g_{rr} = 1 \, , \quad 
g_{\vartheta \vartheta} = r^2 \, .
\end{equation}
We choose the plasma density
\begin{equation}
\omega _p  (r , \vartheta ) ^2
=
\omega _c^2
\,  \mathrm{sin}^2 
 \dfrac{r}{r_0} \, ,
\label{eq:Minkomegap}
\end{equation}
where $\omega _c$ is a constant with the dimension of a 
frequency and $r_0$ is a constant with the dimension of a 
length. Again, $- \omega _p ^2 \, g_{tt} $ is bounded on
the half-plane $\mathcal{U}$ with $\mathrm{sup}
\big( - \omega _p ^2 \, g_{tt} \big) = \omega _c^2$. 
We choose an $\omega _0 > \omega _c$ which guarantees 
that 
\begin{equation}
\mathcal{R}{}_+ ^{\varepsilon} ( r , \vartheta ) = 
\mathcal{R}{}_-  ^{\varepsilon} ( r , \vartheta ) =
- \, r \, \mathrm{sin} \, \vartheta \, 
\sqrt{\omega _0^2 -  \varepsilon \, 
\omega _c ^2 \, \mathrm{sin}{}^2
\dfrac{r}{r_0}}
\label{eq:RMinkowski}
\end{equation}
is defined on the entire half-plane $r \, \mathrm{sin} \, \vartheta > 0$,
for all $\varepsilon \in [0,1]$.
In this case, $\mathcal{R}{}_+ ^1$ and $\mathcal{R}{}_+ ^0$
are, indeed, Morse functions. However, as the gradient of 
$\mathcal{R}{}_+ ^1$ is 
\begin{equation}
\dfrac{
\partial \mathcal{R}{}_{\pm}^1 ( r , \vartheta )
}{
\partial r
} 
=
- \, \dfrac{
\mathrm{sin} \, \vartheta
\Bigg( 
\omega _0 ^2 - 
 \omega _c ^2 \mathrm{sin} \, \dfrac{r}{r_0}
\Big(\mathrm{sin} \, \dfrac{r}{r_0} + \dfrac{r}{r_0} \, 
\mathrm{cos} \, \dfrac{r}{r_0} \Big) \Bigg)
}{
\sqrt{\omega _0^2 -
 \omega _c ^2 \, \mathrm{cos}{}^2
\dfrac{r}{r_0}}
}
\, ,
\end{equation}
\begin{equation}
\dfrac{
\partial \mathcal{R}{}_{\pm}^1 ( r , \vartheta )
}{
\partial \vartheta
} 
=
- \, r \,
\mathrm{cos} \, \vartheta
\, 
\sqrt{\omega _0^2 -
\omega _c ^2 \, \mathrm{cos}{}^2
\dfrac{r}{r_0}} \, ,
\end{equation}
$\mathcal{R}{}_{\pm}^1$
has infinitely many isolated critical points, located at 
\begin{equation}
\omega _0 ^2 = \omega _c^2 \, \mathrm{sin} \, \dfrac{r}{r_0}
\Big(\mathrm{sin} \, \dfrac{r}{r_0}
+ \dfrac{r}{r_0} \, 
\mathrm{cos} \, \dfrac{r}{r_0} \Big) 
\, , \quad
\vartheta = \dfrac{\pi}{2} \, , 
\end{equation}
so Proposition \ref{prop:Brouwer} is again not applicable; we cannot find a 
compact subset of $\mathcal{U}$ which contains all critical
points of $\mathcal{R}{}_{\pm}^1$. This can happen only
because (\ref{eq:infgrad}) is violated for $r \to \infty$.
Fig. \ref{fig:Minkowski1} shows the potential (\ref{eq:RMinkowski}) 
for $\omega _0 = \sqrt{4/3} \,  \omega _c$. 

\begin{figure}
\centerline{\includegraphics[width=0.6 \textwidth]{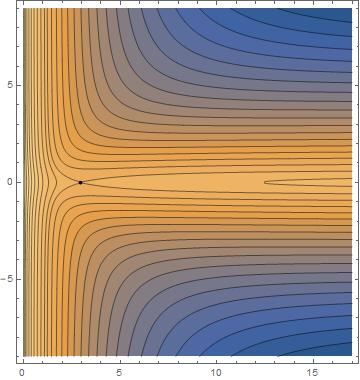}}
\caption{
This picture shows the equipotential lines 
$\mathcal{R}{}_+ = \mathcal{R}{}_- = \mathrm{constant}$ 
for the plasma density (\ref{eq:Monkomegap3}) on Minkowski
spacetime in the $(\rho , z )$-half-plane $\mathcal{U}$, with 
$\rho _0$ used as the units on the axes. We have chosen  
$\omega _0$ equal to $\omega _c$ which implies that 
the strict inequality $\omega _0 > \omega _p (\rho , z )$
holds for all $0 < \rho < \infty$ but not in the limit 
$\rho \to \infty$. The picture shows that there is exactly 
one circular light ray, namely a saddle, marked by a black dot.
In this case the assumptions of Proposition \ref{prop:Brouwer}
are violated because it is impossible to find a one-parameter
family of compact sets such that the critical points of
$\mathcal{R}{}_{\pm}^{\varepsilon}$ are in their interior
for all $\varepsilon \in [0, 1]$. 
} 
\label{fig:Minkowski3}
\end{figure}

For the final example on Minkowski spacetime we work
again in cylindrical polar coordinates (\ref{eq:Minkowski}).
We choose the plasma density
\begin{equation}
\omega _p (\rho , z ) ^2= 
\dfrac{
\omega _c^2(\rho _0^2 + (\rho - \rho _0)^2 )
}{
2 r_0 ^2+ (\rho -\rho_0)^2 + z^2
}
\label{eq:Monkomegap3}
\end{equation}
where again $\omega _c$ is a constant with the dimension of
a frequency and $\rho _0$ is a constant with the dimension of a
length. This time we choose $\omega _0 = \omega _c$ for
which the potentials 
\begin{equation}
\mathcal{R}{}_+^{\varepsilon} ( \rho , z )
=
\mathcal{R}{}_-^{\varepsilon} ( \rho , z )
=
\rho \, \sqrt{\omega _0 ^2 - 
\dfrac{\varepsilon \, \omega _c^2 (r_0^2 + (\rho - r_0)^2 )
}{
2 r_0 ^2+ (\rho -r_0)^2 + z^2
}
}
\end{equation}
are indeed defined on the entire half-plane $\big\{ (\rho , z ) \, \big| \, 
\rho >0 \, , \, - \infty < z < \infty \big\}$. In this case
$\mathcal{R}{}_{\pm}^1$ and $\mathcal{R}{}_{\pm}^0$
are Morse functions.  $\mathcal{R}{}_{\pm}^0$ has no critical 
points whereas $\mathcal{R}{}_{\pm}^1$ has exactly one
critical point, namely a saddle, see Fig. \ref{fig:Minkowski3}. 
Clearly, this can be reconciled with Proposition \ref{prop:Brouwer} 
only if one of the assumptions of this Proposition is violated. This
is indeed the case: For $0.992 \lessapprox \varepsilon < 1$ the 
potential $\mathcal{R}{}_{\pm}^{\varepsilon}$ has \emph{two}
critical points, a saddle and a local maximum. With $\varepsilon
\to 1$ the maximum moves to infinity while the saddle approaches
the saddle of the potential $\mathcal{R}{}_{\pm}^1$ which 
lies in the interior of the half-plane $\mathcal{U}$. Therefore it 
is impossible to include all critical points in a compact set, or in 
a family of compact sets that depend continuously on $\varepsilon$. 
This situation cannot occur if $\omega _0^2$ is \emph{strictly}
bigger than the supremum of $- \omega _p^2 \, g_{tt} $. 

\section{Black-hole spacetimes}
\label{sec:blackh}

In this section we apply Proposition \ref{prop:Brouwer} to
the domain of outer communication $\mathcal{U}$
of a black hole. We assume that $g_{\varphi \varphi} > 0$,
i.e. that there is no causality violation, on $\mathcal{U}$
which guarantees that $t$ is a time function. 
Moreover, we restrict to the case that the spacetime 
is asymptotically flat and that the axis is regular outside 
the horizon, recall Sections \ref{subsec:ax} 
and \ref{subsec:inf}. Note that these assumptions
are \emph{not} satisfied, e.g.,  in the NUT spacetime,
which features a causality-violating region. 

The domain of outer communication is the set
\[
\mathcal{U} = 
\{ ( \rho, z )
\, | \, r_h^2 < \rho ^2 + z^2 \, , \, \rho > 0 \, , \, - \infty < z < \infty \}
\]
\begin{equation}
\quad =
\{ ( r \, \mathrm{sin} \, \vartheta , r \, \mathrm{cos} \, \vartheta )
\, | \, r_h < r < \infty \, , \, 0 < \vartheta < \pi  \}
\end{equation}
where $r = r_h$ is the horizon. 

We first consider black holes without an ergoregion, i.e.,
we assume that $g_{tt} < 0$ on all of $\mathcal{U}$. 
If we connect with the vacuum case by the one-parameter
family of potentials (\ref{eq:Repsilon}), our assumptions
guarantee that $\mathcal{R}_{\pm}^{\varepsilon}$
is defined on $\mathcal{V}{}_{\pm} = \mathcal{U}$,
for all $\varepsilon \in [0,1]$, provided that 
$ - \omega _p^2 \, g_{tt} $ is bounded on $\mathcal{U}$
and that we choose a frequency
constant $\omega _0$ with $\omega _0 ^2 > 
\mathrm{sup} \big( - \omega _p^2 \, g_{tt} \big)$
where the supremum is to be taken over $\mathcal{U}$.
Moreover, if the plasma 
density satisfies conditions (\ref{eq:horgrad}), 
(\ref{eq:axgrad}) and (\ref{eq:infgrad}) at the 
horizon, at the regular axis and at infinity, respectively,
then the results of Sections \ref{subsec:hor}, 
\ref{subsec:ax} and \ref{subsec:inf} guarantee that there
is a compact subset $\mathcal{K}$ of $\mathcal{U}$
such that the citical points of 
$\mathcal{R}{}_{\pm}^{\varepsilon}$ lie in the interior
of $\mathcal{K}$, for all $\varepsilon \in [0, 1]$. 
This compact set is of the form
\begin{equation}
\mathcal{K} =
\big\{ (r \, \mathrm{sin} \, \vartheta , r \, \mathrm{cos} \, \vartheta )
\, \big| \, r \, \mathrm{sin} \, \vartheta \ge  \rho_0 \, , \quad
r_h + \delta \le r \le r_0 \big\}
\end{equation}
with some positive $\rho _0$, $\delta$ and $r_0$. 
Then the one-parameter family (\ref{eq:Repsilon}) satisfies the 
assumptions of Proposition \ref{prop:Brouwer} for all
$\varepsilon \in [0 , 1]$, provided
that $\mathcal{R}{}_{\pm}^1 = \mathcal{R}{}_{\pm}$
and $\mathcal{R}{}_{\pm}^0 $ are Morse functions. 
In this case, the number of critical 
points of $\mathcal{R}{}_{\pm}^1$ differs from the number
of critical points of $\mathcal{R}{}_{\pm}^0$ by an even
number. As $\mathcal{R}{}_{\pm}^{\varepsilon} \to 
- \infty$ at the horizon and at infinity, each 
$\mathcal{R}{}_{\pm}^{\varepsilon}$ must have at least 
one critical point. Actually, for $\mathcal{R}{}_{\pm}^0$
(i.e., vacuum light rays) in 
spherically symmetric and static black-hole spacetimes
the latter observation is rather trivial and well known, see 
e.g. Hasse and Perlick \cite{HassePerlick2002}, Section 6.1,
or Perlick \cite{Perlick2004}, Section 4.3.

We now turn to the case that there is an ergoregion. Disregarding
more complicated situations, we will assume in the following 
that $\mathcal{U}^{\mathrm{out}}$, 
$\mathcal{U}{}^{\mathrm{erg}}$ and $\mathcal{S}$
are connected, where we 
use the notation of Section \ref{subsec:dom}. To fix the
sign ambiguity of the potentials inside an ergoregion (recall
Section \ref{subsec:sign}), we assume that there 
$g_{t \varphi} < 0$.
This is no restriction of generality, because we are free to
transform $\varphi \mapsto - \varphi$ and $g_{t \varphi}$
cannot have zeros inside an ergoregion. If the ergoregion
extends to the horizon, this sign convention
means that prograde ($\dot{\varphi} > 0$) light rays are
co-rotating with the horizon. 

The potential $\mathcal{R}{}_+$, which describes future-oriented 
($ \dot{t} > 0$) prograde ($ \dot{ \varphi} > 0$) light rays, 
is defined on $\mathcal{V}{}_+ = \mathcal{U}$. In contrast 
to the case without an ergoregion, it is not guaranteed that 
there is at least one critical point, because $\mathcal{R}{}_+$ 
does not go to $- \infty$ at the horizon. If the horizon is 
non-degenerate, our assumptions on the plasma density 
imply that the critical points of $\mathcal{R}{}_+$ lie inside 
a compact set. If $\mathcal{R}{}_+ ^1$ and $\mathcal{R}{}_+^0$
are Morse functions, Proposition \ref{prop:Brouwer} guarantees
that the number of critical points of $\mathcal{R}{}_+ ^1$ differs
from the number of critical points of $\mathcal{R}{}_+^0$
by an even number.

The potential $\mathcal{R}{}_-$
has to be considered on $\mathcal{U}{}^{\mathrm{out}}$
and $\mathcal{U}{}^{\mathrm{erg}}$ separately. If 
$ - \omega _p ^2 \, g_{tt} $ is bounded on 
$\mathcal{U}{}^{\mathrm{out}}$, we can choose 
a frequency constant $\omega _0$ with 
$\omega _0^2 > \mathrm{sup} \big( - \omega _p ^2 \, g_{tt} \big)$,
where the supremum is to be taken over $\mathcal{U}{}^{\mathrm{out}}$.
Then $\mathcal{R}{}_-$ is defined on $\mathcal{V}{}_{-} =
\mathcal{U}{}^{\mathrm{out}}$.  On this domain $\mathcal{R}{}_-$ 
describes future-oriented ($\dot{t} > 0$) retrograde ($ \dot{\varphi} < 0$) 
light rays. $\mathcal{V}{}_{-} = \mathcal{U}{}^{\mathrm{out}}$ extends 
to infinity and its boundary may contain points on the regular axis, on the
set $\mathcal{S}$ where $g_{tt} = 0$ and on the horizon. In order to make
sure that all critical points of $\mathcal{R}{}_-$ in 
$\mathcal{U}{}_{\mathrm{out}}$ lie inside a compact set, 
we need the additional assumptions that the horizon is non-degenerate
(if   $\mathcal{U}{}_{\mathrm{out}}$ extends to the horizon) and that 
$\partial g_{tt} / \partial r$ does not go to zero  if a point on 
$\mathcal{S}$ is approached from $\mathcal{U}{}^{\mathrm{out}}$.
If $\mathcal{R}{}_-^1$ and $\mathcal{R}{}_-^0$ are Morse 
functions, Proposition \ref{prop:Brouwer} guarantees that the number
of critical points of $\mathcal{R}{}_-^1$ differ from the number of 
critical points of $\mathcal{R}{}_-^0$ by an even number. As
$\mathcal{R}{}_-^{\varepsilon}$ goes to $- \infty$ for $r \to \infty$
and if a point on $\mathcal{S}$ is approached, this potential must
have at least one critical point, for all $\varepsilon \in [0,1]$.

On the other hand, we 
can also consider $\mathcal{R}{}_-$ on the domain $\mathcal{V}{}_{-}=
\mathcal{U}{}^{\mathrm{erg}}$. There the potential $\mathcal{R}{}_-$ 
is defined for all $\omega _0 > 0$ and if the supremum of $- \omega _p ^2 
\, g_{tt}$ is strictly negative on $\mathcal{U}{}^{\mathrm{erg}}$ also for
$\omega _0 = 0$.  On this domain $\mathcal{R}{}_-$ describes past-oriented
($\dot{t} < 0$) retrograde ($ \dot{\varphi} < 0$) light rays, 
i.e, if we reparametrise the light rays in the
future-oriented sense, then they are prograde. This reflects the known
fact that inside an ergoregion timelike or lightlike curves cannot be
retrograde. (Recall that we have fixed the sign ambiguity by 
requiring $g_{t \varphi} < 0$ in the ergoregion.) The boundary  
of $\mathcal{U}{}^{\mathrm{erg}}$ may consist of points on the set 
$\mathcal{S}$ where $g_{tt} = 0$ and of points on the horizon, 
including possibly the points where the horizon meets the axis. If
$\partial g_{tt} / \partial r$ does not go to zero if a point on
$\mathcal{S}$ is approached from $\mathcal{U}^{\mathrm{erg}}$,
and if the horizon is non-degenerate, our assmptions guarantee 
that the critical points of $\mathcal{R}{}_-$ in 
$\mathcal{U}{}^{\mathrm{erg}}$ lie inside a compact set.
Hence, if $\mathcal{R}{}_-^1$ and $\mathcal{R}_-^0$ are
Morse functions, the number of critical points of  
$\mathcal{R}{}_-^1$ differs from the number of critical
points of $\mathcal{R}_-^0$ by an even number.   
In this case the existence of at least one critical point is not 
guaranteed because $\mathcal{R}{}_-$ approaches a finite value
at the horizon.  
 
An important conclusion from this section is  that in a black-hole spacetime
with an ergoregion that extends to the horizon the existence of a future-oriented 
co-rotating circular light ray is not guaranteed, while there is always at least 
one counter-rotating circular light ray, necessarily outside the ergoregion. So 
it is possible that in the vacuum case there is an even number of co-rotating 
circular light rays which are all destroyed if a plasma is introduced. 


\subsection{Schwarzschild spacetime}

In spherical polar coordinates $(t , \varphi , r , \vartheta )$ the
Schwarzschild spacetime has the following non-zero metric coefficients:
\begin{equation}
g_{tt} (r , \vartheta )= - \Big( 1 - \dfrac{2m}{r} \Big) \, , \quad
g_{\varphi \varphi} (r , \vartheta ) = r^2 \mathrm{sin} \, \vartheta \, ,
\end{equation}
\begin{equation}
g_{rr} (r , \vartheta ) = \Big( 1 - \dfrac{2m}{r} \Big) ^{-1} \, , \quad
g_{\vartheta \vartheta} (r , \vartheta ) = r^2 
\end{equation}
with a positive mass parameter $m$, hence
\begin{equation}
\mathcal{R}_+ (r , \vartheta)= \mathcal{R}_- ( r , \vartheta ) =
- \dfrac{r ^2 \mathrm{sin} \, \vartheta}{r-2m}
\, 
\sqrt{\omega _0 ^2 - \omega _p (r, \vartheta ) ^2 \Big( 1 - \dfrac{2m}{r} \Big)}
\, .
\label{eq:RSchwarzschild}
\end{equation}
(\ref{eq:det2}) is valid on the domain of outer communication
\begin{equation}
\mathcal{U} =
\big\{ ( r \, \mathrm{sin} \, \vartheta , r \, \mathrm{cos} \, \vartheta ) \,
\big| \, r > 2m \, , \, 0 <  \vartheta < \pi \big\} \, .
\end{equation}
If the plasma density is a function of $r$ only, we may restrict to the 
equatorial plane and discuss the properties of light rays in terms of 
a one-dimensional effective potential $V_{\mathrm{eff}} (r)$. 
So if we want to demonstrate the merits of the formalism developed
here we have to consider plasma densities that depend on $r$ and 
$\vartheta$. We will assume that $- \omega _p^2 \, g_{tt}$ 
is bounded on $\mathcal{U}$. Then the potential 
(\ref{eq:RSchwarzschild}) is defined on all of $\mathcal{U}$ for 
$\omega _0 ^2 > \mathrm{sup} \big( - \omega _p^2 \, g_{tt} \big)$.

In vacuum, calculating the gradient of $\mathcal{R}_+ =
\mathcal{R}_-$ and 
equating it to zero produces the well-known result that 
there is one circular light ray, located at $r = 3m$ and 
$\vartheta = \pi /2$. Calculating the second derivatives
shows that this critical point is a maximum (unstable) in 
the $r$-direction and a minimum (stable) in the $\vartheta$ 
direction, so it is a saddle.  

In the plasma case, with $\omega _0 ^2 > 
\mathrm{sup} \big( - \omega _p^2 \, g_{tt} \big)$, 
we read from (\ref{eq:RSchwarzschild}) that $\mathcal{R}_+
= \mathcal{R}_-$ 
goes  to $- \infty$ at infinity and at the horizon, which 
exemplifies our general results. This observation already 
demonstrates that there must be at least  one circular light 
ray in $\mathcal{U}$.

Stronger statements are possible if we use Proposition \ref{prop:Brouwer}.
If (\ref{eq:horgrad}), (\ref{eq:axgrad}) and (\ref{eq:infgrad}) are 
satisfied if the horizon, the regular axis or infinity is approached, the 
critical points of the potential $\mathcal{R}{}_{\pm}^{\varepsilon}$ 
are confined to a compact set. If $\mathcal{R}{}_{\pm} ^ 1$ is a 
Morse function, Proposition \ref{prop:Brouwer}  shows that
in comparison to the vacuum case (one saddle) there can be $2n$
additional circular light rays, $n$ extrema and $n$ saddles,
so  there are $n$ extrema and $n+1$ saddles.

If the plasma density is symmetric with respect to the equatorial
plane, then each extremum in the northern hemisphere must be
accompanied by a similar extremum in the southern hemisphere,
and the same is true for saddles. Therefore, at least one extremum 
lies in the equatorial plane if $n$ is odd and at least one saddle 
lies in the equatorial plane if $n$ is even.

As a specific example, we consider the plasma density.
\begin{equation}
\omega _p^2
=
\dfrac{\omega _c^2 \, 
\big( r^2 + 16 \, m^2 \, \mathrm{sin} ^2 \vartheta \big)}{r^2} 
\label{eq:Schwomegap}
\end{equation}
where $\omega _c$ is a constant with the dimension of a 
frequency.
It was already briefly mentioned in Perlick \cite{Perlick2023} that in 
this case there are three circular light rays, one in the equatorial plane 
and two off the equatorial plane. For off-equatorial circular light rays 
in a plasma on Schwarzschild or Kerr spacetime we also refer to 
Perlick and Tsupko \cite{PerlickTsupko2024}.

\begin{figure}
$\qquad \qquad \quad $
\includegraphics[width=0.45 \textwidth]{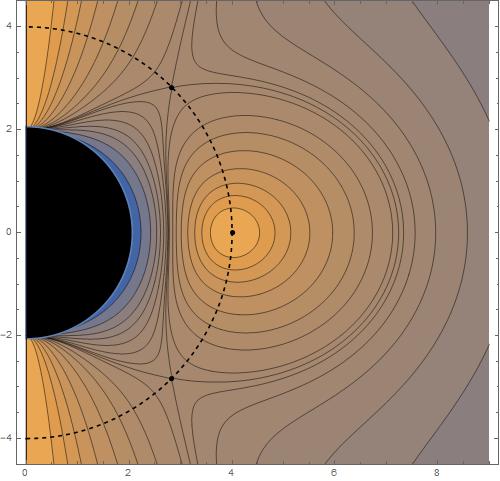}
\hspace{-0.25cm}
\includegraphics[width=0.5 \textwidth]{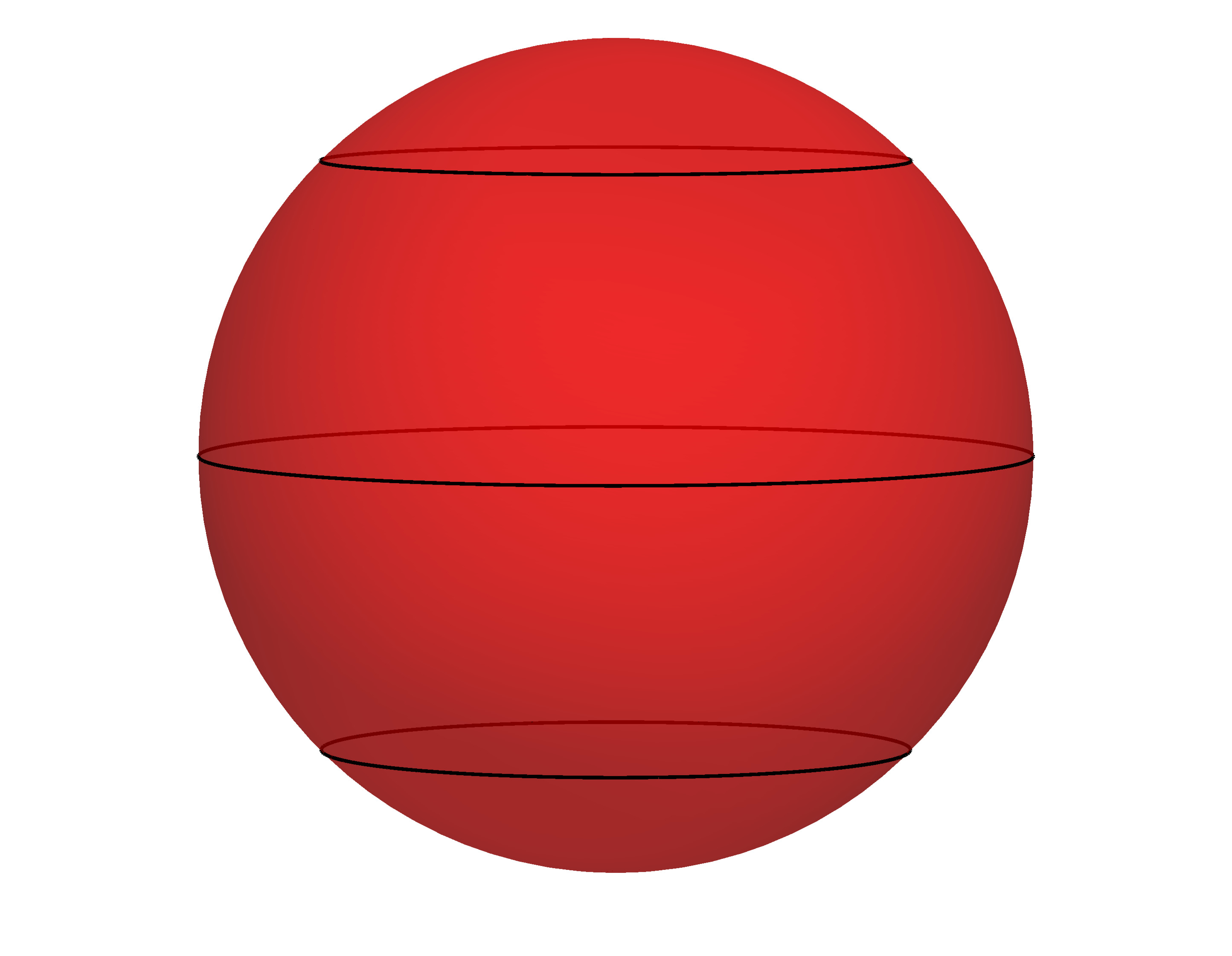}
\caption{
The picture on the left shows the equipotential lines 
$\mathcal{R}{}_+ = \mathcal{R}{}_- = \mathrm{constant}$ 
for the plasma density (\ref{eq:Schwomegap}) on Schwarzschild
spacetime with $\omega _0 = \sqrt{1.001} \, \omega _c$. 
We plot $\rho = r \, \mathrm{sin} \, \vartheta$ on the horizontal axis 
and $z = r \, \mathrm{cos} \, \vartheta$ on the vertical axis,
choosing $m$ as the unit on both axes.
The region inside the horizon is shown as a black disc. The
potential goes to $- \infty$ at the horizon and at infinity, while
it goes to zero on the axis; this confirms our general results
of Sections \ref{subsec:hor}, \ref{subsec:ax} and \ref{subsec:inf}. 
There are three 
circular light rays, marked by black dots: a local maximum 
of the potential in the equatorial plane and two saddles off 
the equatorial plane. They all lie on the circle $r = 4 m$ 
which is marked by a dashed line in the picture on the left. 
In the picture on the right, this sphere with its three circular 
light rays is shown in three-space.  The two off-equatorial 
light rays are unstable in $r$-direction and stable in 
$\vartheta$-direction, while the light ray in the 
equatorial plane is unstable in both directions. Light rays from
the asymptotic region can spiral asymptotically towards these
three circular light rays. This is relevant for the construction of
the shadow in this spacetime which turns out to be a circular
disc, cf. Perlick and Tsupko \cite{PerlickTsupko2017}, 
Section VII.
} 
\label{fig:Schwarzschild}
\end{figure}


\subsection{Kerr spacetime}

In the Kerr spacetime with mass parameter $m$ and spin
parameter $s$, the non-zero metric cofficients in
Boyer-Lindquist coordinates are
\[
g_{tt} = - \dfrac{
r^2+ a^2 \mathrm{cos}{}^2 \vartheta -2mr
}{
r^2+ a^2 \mathrm{cos}{}^2 \vartheta
}
\, , \quad
g_{t\varphi} = 
-\dfrac{
2 \, m \, a \, r \, \mathrm{sin} ^2 \vartheta
}{
r^2+ a^2 \mathrm{cos}{}^2 \vartheta
}
\, , 
\]
\begin{equation}
g_{\varphi \varphi} = 
\mathrm{sin} ^2 \vartheta
\Big(
r^2 + a^2 +
\dfrac{
2 \, m \, r \, a^2\, \mathrm{sin} ^2 \vartheta
}{
r^2+ a^2 \mathrm{cos}{}^2 \vartheta
}
\Big) 
\end{equation}
\begin{equation}
g_{rr} = 
\dfrac{
r^2+ a^2 \mathrm{cos} ^2 \vartheta
}{
r^2-2mr+a^2
}
\, , \quad
g_{\vartheta \vartheta} = 
r^2+ a^2 \mathrm{cos} ^2 \vartheta
\, ,
\end{equation}
hence
\[
\hspace{-1.75cm}
\mathcal{R}{}_{\pm}
=
\pm \dfrac{
2 \, \omega _0 \, m \, a \, r \, \mathrm{sin} ^2 \vartheta
}{
r^2+ a^2 \mathrm{cos}{}^2 \vartheta -2mr
}
\]
\begin{equation}
\hspace{-1.75cm}
-
\dfrac{
r^2+ a^2 \mathrm{cos}{}^2 \vartheta
}{
r^2+ a^2 \mathrm{cos}{}^2 \vartheta -2mr
}
\sqrt{r^2-2mr+a^2} \; \mathrm{sin} \, \vartheta
\, 
\sqrt{
\omega _0 ^2 - \omega _p^2
\dfrac{
(r^2+ a^2 \mathrm{cos}{}^2 \vartheta -2mr )
}{
(r^2+ a^2 \mathrm{cos}{}^2 \vartheta )
}
}
\, .
\end{equation}
We consider a Kerr black hole with $0 < a < m$, but we will later
also give an example for the extreme case $a = m$. We concentrate
on the domain of outer communication
\begin{equation}
\mathcal{U} = \big\{ 
(r \, \mathrm{sin} \, \vartheta , r \, \mathrm{cos} \, \vartheta )
\big| \, r > m + \sqrt{m^2-a^2} \, , \: 0 < \vartheta < \pi \big\}
\, .
\end{equation}
 Our general results imply that, if $\mathcal{R}{}_{+}$ and 
$\mathcal{R}{}_-$ are Morse functions and if $\omega _p^2$ 
satisfies the required boundedness conditions, then there is an 
odd number of co-rotating light rays in $\mathcal{U}$ and 
an odd number of counter-rotating light rays in 
$\mathcal{U}^{\mathrm{out}}$, i.e. outside the ergoregion.

\begin{figure}
$\qquad \qquad \quad $ 
\includegraphics[width=0.42 \textwidth]{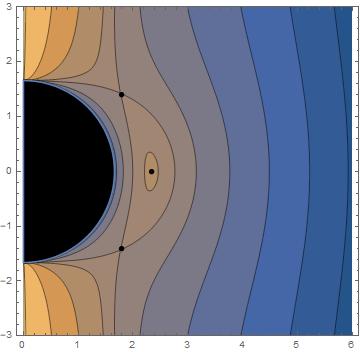}
\hspace{0.25cm}
\includegraphics[width=0.42 \textwidth]{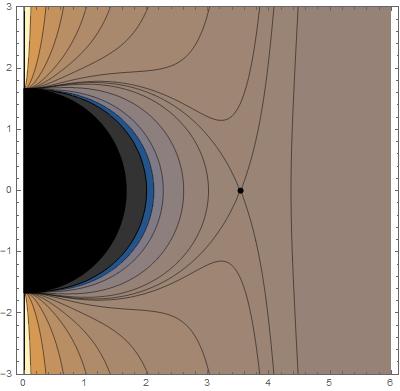}
\caption{These pictures show the equipotential lines
$\mathcal{R}{}_+ = \mathrm{constant}$ (on the left)
and $\mathcal{R}{}_- = \mathrm{constant}$ (on the right) 
for the plasma density (\ref{eq:Kerr1omegap}) on Kerr
spacetime with $a = 3m/4$ and 
$\omega _0 = \sqrt{1.001} \,\omega _c$. On the 
horizontal axis we plot $\rho = r \, \mathrm{sin} \, \vartheta$
and on the vertical axis we plot $z = r \, \mathrm{cos} \, \vartheta$,
using $m$ as the unit on both axes. 
The region inside the horizon is shown as a black disc, in 
the picture on the right the ergoregion is shown in grey. 
There are three corotating light rays in the domain of outer 
communication, marked by black dots in the picture 
on the left, a local maximum of the potential in the 
equatorial plane and two saddles off the equatorial plane. 
The saddles are maxima (unstable) in the $r$ direction and 
minima (stable) in the $\vartheta$ direction. This is similar 
to our Schwarzschild example shown in Fig. \ref{fig:Schwarzschild}.
For the potential $\mathcal{R}{}_-$ we show in the
picture on the right the equipotential lines only on 
$\mathcal{U}{}^{\mathrm{out}}$. There are no
circular light rays for this particular plasma density in
the ergoregion, so we left $\mathcal{U}{}^{\mathrm{erg}}$
in the picture grey. Outside of the ergoregion, we read from 
the picture that there is exactly one circular light ray, and
that it is a saddle, very similar to the vacuum case.
} 
\label{fig:Kerr1}
\end{figure}

As an example we consider  on a Kerr spacetime with 
$a = 3 m /4$ the plasma density
\begin{equation}
\omega _p^2 =
\dfrac{
\omega _c^2  m^2 \big( \mathrm{sin} ^2 (50 \,  r/m)
+ 2 \, \mathrm{sin} ^2 \vartheta \big)
}{
r^2+ a^2 \mathrm{cos}^2 \vartheta
}
\label{eq:Kerr1omegap}
\end{equation}
where $\omega _c$ is a constant with the dimension of a 
frequency. For $\omega _0 > \omega _c$,
$\mathcal{R}{}_+$ is defined on $\mathcal{U}$
while $\mathcal{R}{}_-$ is defined on the disjoint open
sets $\mathcal{U}{}^{\mathrm{out}}$ and
$\mathcal{U}{}^{\mathrm{erg}}$. Then all 
assumptions of Proposition \ref{prop:Brouwer} are
satisfied, so we know that there must be an odd number
of co-rotating circular light rays in $\mathcal{U}$ and 
an odd number of counter-rotating circular light rays
in $\mathcal{U}{}^{\mathrm{out}}$. This is indeed 
the case, as Fig. \ref{fig:Kerr1} demonstrates.

\begin{figure}
\centerline{
\includegraphics[width=0.475 \textwidth]{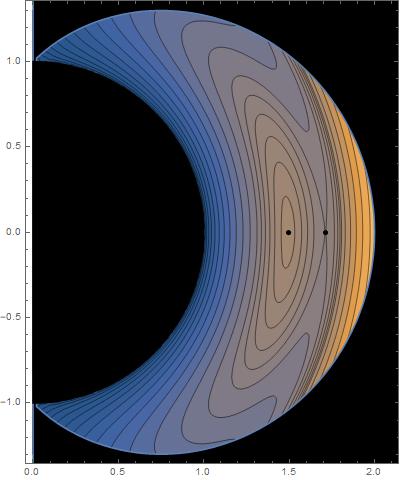}
}
\caption{
This picture shows the equipotential lines 
$\mathcal{R}{}_- = \mathrm{constant}$ for
the plasma density (\ref{eq:Kerr2omegap}) on Kerr
spacetime with $a = m$ and $\omega _0 = 0$. 
Again, we plot $\rho = r \, \mathrm{sin} \, \vartheta$
on the horizontal axis and $z = r \, \mathrm{cos} \,
\vartheta$ on the vertical axis, choosing $m$ as the
unit on both axes.
The regions below the (degenerate) horizon and 
outside of the ergoregion are shown in black, i.e., 
only the ergoregion is displayed. 
There are two circular light rays in the ergoregion, 
marked by black dots in the picture. The inner one
is a local maximum, the outer one is a saddle which
is a minimum  (stable) in $r$ direction and a maximum
(unstable) in the $\vartheta$ direction. Note that in
comparison to the vacuum light rays in (Schwarzschild 
or) Kerr spacetime the stability properties of the
saddle are reversed. In this case the assumptions
of Proposition \ref{prop:Brouwer} are indeed satisfied,
but this was not implied by our general results because
the horizon is degenerate.
} 
\label{fig:Kerr3}
\end{figure}

We have mentioned several times that in a plasma circular
light rays with $\omega _0 = 0$ are possible in an ergoregion.
Here is an example that illustrates this fact. We choose
an extremal Kerr black hole, $a = m$, and a plasma density
\begin{equation}
\omega _p^2 =
\omega_c ^2 
\Bigg( \dfrac{3}{2}  + \mathrm{sin} ^2 \Big( \dfrac{2 \pi r}{m} \Big)
 \Bigg)
\label{eq:Kerr2omegap}
\end{equation}
where $\omega _c$ is a again a constant with the dimension of
a frequency. Inside the ergoregion, the potential $\mathcal{R}{}_-$ 
is defined for $\omega _0 = 0$. There are two counter-rotating
circular light rays in this domain, see Fig. \ref{fig:Kerr3}.

\subsection{NUT spacetime}

The NUT metric, which was found by Newman, Unti and
Tamburino \cite{NewmanTamburinoUnti1963} in 1963 as
a solution to Einstein's vacuum field equation, depends
on two parameters which have the dimension of a length,
a mass parameter $m$ and a NUT parameter, also known
as gravitomagnetic charge, $n$. For a detailed discussion
we refer to Griffiths and Podolsk{\' y} \cite{GriffithsPodolsky2009}.

The non-zero metric coefficients are
\[
g_{tt} = - \dfrac{r^2-2mr-n^2}{r^2+n^2}
\, , \quad
g_{t\varphi} = 
\dfrac{
(r^2-2mr-n^2) \, 2 \, n \, \mathrm{cos} \, \vartheta
}{
r^2+n^2
}
\, , 
\]
\begin{equation}
g_{\varphi \varphi} = 
(r^2+n^2) \, \mathrm{sin} ^2 \vartheta
-
\dfrac{
(r^2-2mr-n^2) \, 4 \, n^2 \, \mathrm{cos} ^2 \vartheta
}{
r^2+n^2
}
\end{equation}
\begin{equation}
g_{rr} = \dfrac{r^2+n^2}{r^2-2mr-n^2}
\, ,  \quad
g_{\vartheta \vartheta} = r^2+n^2 \, , 
\end{equation}
hence
\[
\mathcal{R}{}_{\pm} = 
\pm
\dfrac{r^2-2mr-n^2}{r^2+n^2}
\, 2 \, n \, \mathrm{cos} \, \vartheta
\, \omega _0
\]
\begin{equation}
+ 
\dfrac{\sqrt{r^2-2mr-n^2}}{\sqrt{r^2+n^2}}
 \, \mathrm{sin} \, \vartheta \,
\sqrt{ \omega _0^2  (r^2-2mr-n^2)- \omega _p^2 
(r^2+n^2)} \, .
\end{equation}
The NUT metric features a black-hole horizon at 
$r_h = m + \sqrt{m^2+n^2}$. In the following
we concentrate on the domain of outer communication
\begin{equation}
\mathcal{U} = \big\{
(r \, \mathrm{sin} \, \vartheta , r \, \mathrm{cos} \, \vartheta )
\big| \, r > m + \sqrt{m^2+n^2} \, , \: 0 < \vartheta < \pi \, 
\big\}
\, .
\end{equation} 
Manko and Ruiz \cite{MankoRuiz2005} have introduced 
an additional parameter, $C$, into the NUT metric. Here
we have chosen $C =0$. With this choice, the NUT metric
is singular on both axes, $\vartheta =0$ and $\vartheta
= \pi$, outside the horizon and there is a causality-violating
region near these axes. There is no ergoregion, and 
(\ref{eq:det}) is satisfied on all of $\mathcal{U}$.  
Following Bonnor \cite{Bonnor1969}, one may interpret
the singular axes as spinning rods. 

\begin{figure}
\centerline{	\includegraphics[width=0.6 \textwidth]{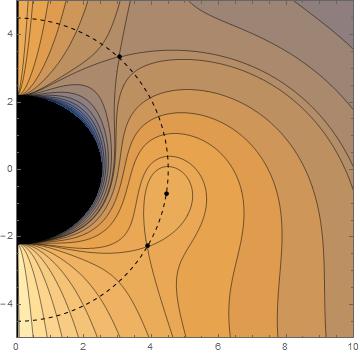}}
\caption{
This picture shows the equipotential lines 
$\mathcal{R}{}_+ = \mathrm{constant}$
for the plasma density (\ref{eq:NUTomegap}) on the
NUT spacetime with $n = 0.635 \, m$
and $\omega _0 = \sqrt{1.001} \, \omega _c$. Again,
we plot $\rho = r \, \mathrm{sin} \, \vartheta$ on the horizontal
axis and $z = r \, \mathrm{cos} \, \vartheta$ on the vertical
axis, with $m$ as the unit on both axes. The region inside the 
horizon is shown as a black disc. There are three circular light rays,
marked by black dots, a local maximum of the potential and two 
saddles. The saddles are maxima (unstable) in $r$ direction and 
minima (stable) in $\vartheta$ direction. As in the Schwarzschild 
example shown in Fig. \ref{fig:Schwarzschild}, the three circular light
rays lie on a sphere, which is marked by a dashed line. Since 
$\mathcal{R}{}_+ (r, \vartheta ) = \mathcal{R}{}_- (r , \pi - \vartheta )$,
the equipotential lines of $\mathcal{R}{}_-$ are given by the same
picture turned upside down.
} 
\label{fig:NUT}
\end{figure}

In vacuum, there are two circular light rays, one prograde
and one retrograde, off the equatorial plane and symmetric
with respect to it, see Jefremov and Perlick 
\cite{JefremovPerlick2016}. More generally, if 
$\omega _p (r  , \vartheta ) = \omega _p (r , \pi - \vartheta)$,
one potential is the mirror image of the other,  
$\mathcal{R}{}_ +(r , \vartheta ) = 
\mathcal{R}{}_- ( r , \pi - \vartheta )$,
i.e., for every prograde circular light ray in the 
northern hemisphere there is a retrograde one 
in the southern hemisphere and vice versa.

As a specific example we choose 
$n = 0.635 \, m$ and
\begin{equation}
\omega _p ( r , \vartheta )^2 
= 
\omega _c^2 
\, \Big(1 + \dfrac{16 m^2}{r^2} \mathrm{sin}{}^2 \vartheta \Big)
\label{eq:NUTomegap}
\end{equation}
where $\omega _c$ is a constant with the dimension of a frequency.
If $\omega _0 > \omega _c$, the one-parameter family
(\ref{eq:Repsilon}) satisfies all assumptions of Proposition
\ref{prop:Brouwer}, for both signs, so there must be an
odd number of retrograde and an odd number of prograde
circular light rays in the plasma. This is indeed true, as is illustrated 
by Fig. \ref{fig:NUT}.

\noindent
\section{Conclusions}
The potentials $\mathcal{R}_+$ an $\mathcal{R}{}_-$ which are
at the centre of this paper are useful for several reasons. Firstly,
they illustrate the influence of a plasma on the lensing features
in axially symmetric and stationary spacetimes in a suggestive
way. In particular, plotting the corresponding equipotential 
surfaces (i.e., the generalised Von Zeipel cylinders)
immediately locates the circular light rays and allows
to read the direction of the centrifugal-plus-Coriolis force
experienced by light rays. Secondly, and maybe 
even more importantly, these potentials allow to determine in 
a mathematically precise way how the light rays in a plasma 
differ from the light rays in vacuum on the same spacetime. 
This is highly relevant, in particular, in the spacetime of a black 
hole or some other ultracompact object: With the help of 
Proposition \ref{prop:Brouwer} one can find out if the circular 
vacuum light rays in the spacetime persist if a plasma is taken 
into account. Among other things, the answer to this question 
gives important information on the shadow of such an object 
in the presence of a plasma. 

Roughly speaking, Proposition \ref{prop:Brouwer} guarantees
that generically the number of circular light rays in a plasma
differs from the number of circular vacuum light rays on the
same spacetime by an even number, provided that the plasma 
density satisfies certain boundedness conditions. As a 
consequence, if there is an odd number of circular light
rays in vacuum, it is impossible that all of them are
destroyed when a plasma is introduced. More specifically,
we have seen that on the Kerr spacetime Proposition 
\ref{prop:Brouwer}, if applicable, implies that in a plasma
there is at least one co-rotating and at least one 
counter-rotating circular light ray in the domain of outer 
communication, where one counter-rotating circular light 
ray must be outside the ergoregion. Again, this is of 
relevance for the shadow. 

It is to be emphasised that throughout the present study 
Einstein's field equation was not used. Therefore, the general 
results may be applied to the case of a self-gravitating plasma 
and also to the case that the gravitational field of the plasma is 
ignored. In the examples we have restricted ourselves to the 
latter case.

It is certainly desirable to generalise the techniques used in this 
paper beyond the case of circular light rays, namely to spatially 
bounded light rays. Potentials very similar to the ones used
in this paper have been utilised by Hasse and Perlick
\cite{HassePerlick2006} and by Halla and Perlick \cite{HallaPerlick2022}
for characterising those regions in axially symmetric and stationary 
spacetimes where vacuum light rays can have turning points, i.e.,
where the radius coordinate can have
a local maximum or a  local minimum, respectively, along a vacuum 
light ray. However, it is hard to see if and how for that purpose 
the Brouwer degree of the gradient of the potentials can be helpful.

\section*{Acknowledgement}
The idea for this paper grew out of a discussion with Pedro
Cunha during the Workshop on ``Lensing and wave optics 
in strong gravity'' in December 2024  at the Erwin 
Schr{\"o}dinger Institute in Vienna, Austria. Quite generally, 
this workshop proved exceptionally fruitful for me. 

\noindent
\section*{Appendix: Proof of Proposition 1} 
To prove (\ref{eq:pphiR}), we start out from (\ref{eq:Ham1inv}),
\begin{equation}
p_A = g_{AB} \dot{x}{}^B + g_{Ai} \dot{x}{}^i \, ,
\label{eq:pA}
\end{equation}
Our assumption (\ref{eq:det}) guarantees that the matrix
\begin{equation}
\big( g_{AB} \big) = 
\begin{pmatrix}
g_{tt} & g_{t \varphi}
\\
g_{t \varphi} & g_{\varphi \varphi}
\end{pmatrix}
\end{equation}
is invertible. We denote the inverse
\begin{equation}
\big( \gamma ^{AB} \big) =
\dfrac{1}{g_{tt} g_{\varphi \varphi} - g_{t \varphi} ^2}
\, 
\begin{pmatrix}
g_{\varphi \varphi } & - g_{t \varphi}
\\
- g_{t \varphi} & g_{tt}
\end{pmatrix}
\, .
\label{eq:gamma}
\end{equation}
This allows to solve (\ref{eq:pA}) for $\dot{x}{}^C$,
\begin{equation}
\dot{x}{}^C = 
\gamma ^{C B} \big( p_B - g_{bj} \dot{x}{}^j \big)  
\, .
\label{eq:dxC}
\end{equation}
Inserting this expression into (\ref{eq:Ham2}) results in
\[
2 \, \mathcal{H} = 
g_{ij} \dot{x}{}^i \dot{x}{}^j + 2 g_{iA} \dot{x}{}^i
\gamma ^{AB} \big(p_B - g_{Bj} \dot{x}{}^j \big)
\]
\begin{equation}
+
g_{AB} \gamma ^{AC} 
\big(p_B - g_{Bj} \dot{x}{}^j \big)
\gamma ^{BD}
\big(p_D - g_{Dj} \dot{x}{}^j \big)
+ \omega _p^2
\, .
\end{equation}
With $g_{AB} \gamma ^{BC} = \delta _A^C$ this results in
\begin{equation}
\mathcal{H} = \dfrac{1}{2} 
\Big(
F
+
\gamma ^{BD} p_B p_D
+ \omega _p^2
\Big)
\label{eq:Ham4}
\end{equation}
with
\begin{equation}
F = \Big( g_{ij} - g_{iA} \gamma ^{AB} g_{Bj} \Big)
\dot{x}{}^i \dot{x}{}^j
\, .
\label{eq:F}
\end{equation}
We rewrite this equation with the explicit form of the matrix 
$(\gamma ^{AB})$ from (\ref{eq:gamma}):
\begin{equation}
2 \, \mathcal{H} - F = 
\dfrac{
g_{tt} p_{\varphi} ^2 - 2 g_{t \varphi} p_t p_{\varphi}
+ g_{\varphi \varphi} p_t ^2 
}{
g_{tt} g_{\varphi \varphi} - g_{t \varphi} ^2
}
+ \omega _p^2
\, ,
\end{equation}
hence
\begin{equation}
\hspace{-1.6cm}
\dfrac{1}{g_{tt}} \, 
\big(g_{tt}g_{\varphi \varphi} - g_{t \varphi} ^2 \big)
\big( 2 \, \mathcal{H} - F  \big)
= 
p_{\varphi} ^2
 - 2 \dfrac{g_{t \varphi}}{g_{tt}} 
 p_{\varphi} p_t
+ \dfrac{g_{\varphi \varphi}}{g_{tt}} \, p_t^2
+
\big( g_{tt}g_{\varphi \varphi} - g_{t \varphi} ^2 \big)
\, \dfrac{\omega _p^2}{g_{tt}}
\, .
\end{equation}
Factorising the second-order polynomial on the right-hand side yields
\begin{equation}
\hspace{-1cm}
\dfrac{1}{g_{tt}} \, 
\big( g_{tt}g_{\varphi \varphi} - g_{t \varphi} ^2 \big)
\,
\big( 2 \, \mathcal{H} - F  \big)
= 
\Big( p_{\varphi} -
\dfrac{g_{t \varphi}}{g_{tt}} \, p_t - Q \Big)
\Big( p_{\varphi} -
\dfrac{g_{t \varphi}}{g_{tt}} \, p_t + Q \Big)
\label{eq:Ham5}
\end{equation}
where
\begin{equation}
Q = 
\dfrac{1}{g_{tt}
}
\,
\sqrt{g_{t \varphi} ^2 - g_{tt} g_{\varphi \varphi}}
\, 
\sqrt{ p_t^2+ \omega _p^2 g_{tt}}
\label{eq:Q}
\end{equation}
With $\mathcal{H} =0$ and the fact that $F$ vanishes at points
where $\dot{x}{}^i =0$, this proves (\ref{eq:pphiR}). 
To prove (\ref{eq:dotphi}) and (\ref{eq:dott}), we write 
(\ref{eq:dxC}) explicitly at points where $\dot{x}{}^i = 0$:
\begin{equation}
\dot{\varphi} =
\dfrac{
g_{tt} p_{\varphi} - g_{t \varphi} p_t
}{
g_{\varphi \varphi} g_{tt} - g_{t \varphi} ^2
}
\, , \quad
\dot{t} =
\dfrac{
g_{\varphi \varphi} p_t - g_{t \varphi} p_{\varphi}
}{
g_{\varphi \varphi} g_{tt} - g_{t \varphi} ^2
}
\, .
\end{equation}
Inserting (\ref{eq:pphiR}) proves (\ref{eq:dotphi}) and
(\ref{eq:dott}). 
Finally, we prove (\ref{eq:ddx}). To that end we differentiate
(\ref{eq:Ham5}) with respect to $x^i$ and set $\mathcal{H}$ 
and the $\dot{x}{}^k$ equal to zero afterwards. This results in
\begin{equation}
\dfrac{1}{g_{tt}} \, 
\big(g_{tt} g_{\varphi \varphi} - g_{t \varphi}^2 \big)
\, 
\dfrac{\partial \mathcal{H}}{\partial x^i} =
\mp
Q \dfrac{\partial}{\partial x^i} \Big(
- \dfrac{g_{t \varphi}}{g_{tt}} \, p_t \pm Q
\Big) 
\, .
\label{eq:dHam}
\end{equation}
Here we have used the fact that $\partial F / \partial x^i$ vanishes
at points where $\dot{x}{}^i = 0$. This is not quite trivial because
in the Hamiltonian formalism a partial derivative $\partial / 
\partial x^{\mu}$ means that the $p_{\rho}$ are kept fixed,
i.e., we have to express the $\dot{x}{}^i$ in (\ref{eq:F}) in terms of $p_{\rho}$ and $x^{\mu}$ before we can calculate the derivative 
$\partial F/ \partial x^i$. The result gives indeed zero at points 
where $\dot{x}{}^i = 0$, because the $\dot{x}{}^j$ occur 
quadratic in (\ref{eq:F}). 
With (\ref{eq:dHam}) in our hands, we now differentiate
(\ref{eq:Hamx}) which yields
\begin{equation}
\ddot{x}^{\mu} = 
\dfrac{\partial g^{\mu \sigma}}{\partial x^i} 
\dot{x}^i p_{\sigma}
+ g^{\mu i} \dot{p}_i \, .  
\end{equation}
Here we have used that $\dot{p}{}^A = 0$. With the help of (\ref{eq:Hamp}) this can be rewritten at points where  
$\dot{x}{}^i = 0$ as
\begin{equation}
\ddot{x}{}^{\mu} = 
- g^{\mu i} \dfrac{\partial \mathcal{H}}{\partial x^i}
\, .
\end{equation}
Inserting (\ref{eq:dHam}) into this expression proves (\ref{eq:ddx}).
\hfill $\square$

\section*{References}

\bibliographystyle{iopart-num}

\end{document}